\documentclass[preprint]{aastex}
\usepackage{graphicx}
\usepackage{natbib}
\usepackage{epsfig}

\newcommand{\dpi}{\dot{\varpi}_{\rm sec}}           
\newcommand{\ares}{a_{\rm res}}                
\newcommand{\nres}{n_{\rm res}}                

\begin{document}

\title{The Architecture of the Cassini Division}
\author{M.M. Hedman$^*$, P.D. Nicholson}
\affil{Department of Astronomy, Cornell University, Ithaca NY 14853 }
\author{K.H. Baines, B.J. Buratti, C. Sotin}
\affil{Jet Propulsion Lab, Pasadena CA 91109}
\author{R.N. Clark}
\affil{United States Geological Survey, Denver CO 80225}
\author{R.H. Brown}
\affil{Lunar and Planetary Lab, University of Arizona, Tucson AZ 85721}
\author{R.G. French}
\affil{Department of Astronomy, Wellesley College, Wellesley MA 02481}
\author{E.A. Marouf}
\affil{Department of Electrical Engineering, San Jose State University, San Jose CA 95192}

\bigskip

\bigskip

$^*$ Corresponding author, mmhedman@astro.cornell.edu

{\it Proposed Running Head:} Architecture of the Cassini Division

\maketitle

\pagebreak

{\bf ABSTRACT:} The Cassini Division in Saturn's rings contains a series of eight named
gaps, three of which contain dense ringlets. Observations of stellar 
occultations by the Visual and Infrared Mapping Spectrometer onboard 
the Cassini spacecraft have yielded $\sim$40 accurate and precise 
measurements of the radial position of the edges of all of these 
gaps and ringlets. These data reveal suggestive patterns
in the shapes of many of the gap edges: the 
outer edges of the 5 gaps without ringlets are circular to within 1 km,
while the inner edges of 6 of the gaps are eccentric, with 
apsidal precession rates consistent with those expected for eccentric orbits
near each edge. Intriguingly, the pattern speeds of these eccentric 
inner gap edges, together with that of the eccentric Huygens ringlet,
 form a series with a characteristic spacing of 
$0.06^\circ/$day. 

The two gaps with non-eccentric inner edges lie near
first-order Inner Lindblad Resonances (ILRs) with moons. One such edge is 
close to the 5:4 ILR with Prometheus, and the radial excursions
of this edge do appear to have an $m=5$ component aligned
with that moon. The other resonantly confined edge
is the outer edge of the B ring, which lies near the 2:1 Mimas 
ILR. Detailed investigation of the B-ring-edge data confirm
the presence of an $m=2$ perturbation on the B-ring edge, but
also show that 
during the course of the Cassini Mission, this pattern has drifted
backward relative to Mimas. 
Comparisons with earlier occultation measurements going back
to Voyager suggest the possibility that the $m=2$ pattern
is actually librating relative to Mimas with a libration
frequency $L\sim0.06^\circ$/day (or possibly 0.12$^\circ$/day). In addition to the
$m=2$ pattern, the B-ring edge also has an $m=1$ component
that rotates around the planet at a rate close to the expected apsidal 
precession rate ($\dot{\varpi}_B \sim 5.06^\circ/$day). 
Thus the pattern speeds of the eccentric edges in the Cassini
Division can be generated from various combinations
of the pattern speeds of structures observed on the edge of the 
B ring: $\Omega_p=\dot{\varpi}_B-jL$ for $j=1,2,3,...,7.$
We therefore suggest that most of the
gaps in the Cassini Division are produced by resonances
involving perturbations from the massive edge of the B ring.
We find that a combination of gravitational perturbations generated by
the radial excursions in the B-ring edge and the gravitational
perturbations from the Mimas 2:1 Inner Lindblad Resonance yields
terms in the equations of motion that should act to constrain
the pericenter location of particle orbits in the vicinity
of each of the eccentric inner gap edges in the Cassini Division.
This alignment of pericenters could be responsible
for forming the Cassini-Division gaps and thus explain
why these gaps are located where they are.

\noindent {\it Subject Keywords: } occultations, planets: rings

\pagebreak

\section{Introduction}

The Cassini Division is a roughly 4500-km wide region in Saturn's
main rings situated between the A and B rings. Far from being
a completely empty gap between these two rings, this zone is actually
a complex region containing an array of gaps and ringlets. 
The physical processes responsible for creating and maintaining 
most of the observed features in this region remain obscure. 

\begin{figure}[tb]
\resizebox{6in}{!}{\includegraphics{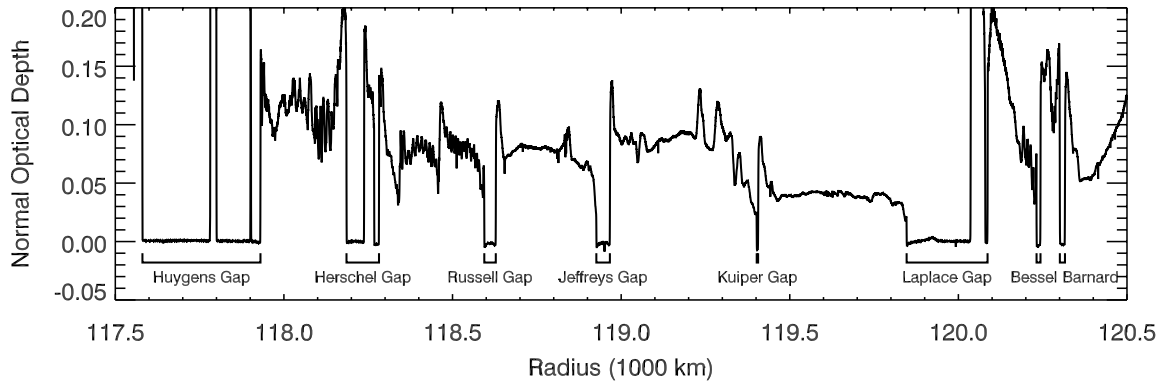}}
\caption{Overview of the basic architecture of the inner part
of the Cassini Division. This optical depth profile is derived
from the Rev 8 ingress occultation of $o$ Ceti by the rings
observed by the VIMS instrument on board Cassini. The various
gaps discussed in this study are labeled.}
\label{cdov}
\end{figure}

In particular, there is still no definitive explanation for most of the 
numerous gaps present throughout the inner part of Cassini
Division. As shown in Figure~\ref{cdov}, there are eight
gaps in the inner part of the Cassini Division, all of which are
now named after various researchers who worked on Saturn's
rings \citep{Colbook}. The innermost gap is called the Huygens Gap and it 
marks the inner boundary of the Cassini Division. The inner edge of 
the Huygens Gap, which is also the outer edge of the 
massive B ring, has long been known to be 
associated with a 2:1 mean-motion resonance with Saturn's moon Mimas \citep{Porco84}. 
This gap contains two  optically thick ringlets: the inner one, 
called the ``Huygens Ringlet", is known to be eccentric, 
while the outer, or ``Strange" ringlet seems to have a 
significant inclination \citep{Turtle91, Spitale06, Spitale08}. 

While the Mimas 2:1 resonance likely plays an important
role in creating the Huygens Gap, the origins of the other 
seven gaps, as well as the dense ringlets in the Huygens, 
Herschel and Laplace gaps, are much less clear.
Some have argued that each gap in the Cassini Division
 contains a tiny moon \citep{Lissauer81}, just as the Encke and Keeler Gaps
 in the A ring are maintained by the small moons Pan and Daphnis.
While some wavelike features in the Cassini Division have been interpreted
as evidence for the existence of such moons~\citep{MT86},  
direct detections of the moons themselves have not yet been reported.
It is also not clear why such moons would be concentrated in this 
particular part of the ring system, well inside Saturn's Roche limit
for icy bodies.

Using the extensive occultation data obtained by the Visual 
and Infrared Mapping Spectrometer (VIMS) onboard the
Cassini spacecraft, we have conducted an 
investigation of the gaps in the Cassini Division. The
high resolution and accurate geometrical information
possible with these data has enabled us to determine
the shapes of most of the edges of these gaps and ringlets. 
Based on this information, we will propose alternative explanations
for these gaps that do not require the existence of a small moon
within each and every gap.

Section 2 of this paper described the VIMS observations used in 
this analysis, and how they were processed to obtain position
estimates for the various edges. Section 3 presents
the results of our analysis and the derived constraints
on the shapes of various edges in the Cassini Division, including
the B-ring edge. The B-ring edge turns out to be quite complex,
so Section 4 briefly compares some of this edge's observed features
to the predictions from simple dynamical models.
Section 5 presents a dynamical model that could explain
the location of the inner edges of the various gaps in the Cassini 
Divsion as the result of a series of resonances involving perturbations from the massive B ring 
edge. This model is only a first step towards understanding the architecture
of the Cassini Division, so Section 6 discusses potential future work 
that could confirm and extend this model. The final section summarizes
our conclusions.

\section{Observations and Data Reduction}

The Visual and Infrared Mapping Spectrometer (VIMS) is most often used to 
produce spatially-resolved spectra of planetary targets between 0.3 and 
5.1 microns. However, VIMS is a flexible instrument that can operate in a 
variety of modes, including an occultation mode~\citep{Brown04}. In this 
mode, the imaging capabilities are disabled, the  short-wavelength 
VIS channel of the instrument is turned off and the IR 
channel obtains a  series of 0.8-5.1 micron spectra from a single pixel 
targeted at a star. Typical  sampling invtervals were 20-80 ms. The data used in this analysis are uncalibrated, but 
a mean instrumental thermal background has been subtracted from all the spectra
for each occultation.
A precise time stamp is appended to every spectrum 
to facilitate reconstruction of the occultation geometry.

\subsection{Observations}

\begin{table}
\caption{Occultation cuts used in this analysis}
\label{obstab}
\centerline{\resizebox{4in}{!}{\begin{tabular}{| c r c c c r c c l |}
\hline
Rev & Star                 & & UTC         & Inertial & max DN  & Radial &  Radial & Notes \\
        &                          & & at Jeffreys Gap & Longitude & at 3 $\mu$m & Sampling (km) & Offset (km) & \\
\hline
08 & $o$Cet	       & i & 2005-144T05:58 & 1$^\circ$ & 995 & 0.30 & +0.06 & A\\
08 & $o$Cet            & e & 2005-144T07:07 &-28$^\circ$ &  999 & 0.30 & -0.68 & A\\
13 & $\alpha$Sco   & i & 2005-232T11:46 &  -70$^\circ$ & 860 & 0.37 & +0.77 & B,I\\
13 & $\alpha$Sco   & e & 2005-232T13:44 & -3$^\circ$ & 770 & 0.38 & -0.98 &I \\
\hline
26 & $\alpha$Ori    & i & 2005-204T16:47 &  -3$^\circ$ & 967 & 0.21 & -0.32 & \\
28 & $\alpha$Tau   & i & 2006-252T10:54 &  34$^\circ$ & 144 & 0.52 & +0.15 & F\\
29 & $\delta$Vir      & i & 2006-268T22:37 & 204$^\circ$ & 125 & 1.44 & -1.35 & Xr,A,C\\
29 & $\delta$Vir      & e & 2006-268T22:44 & 97$^\circ$ & 128 &1.43 & +2.74 & Xr,A,C\\
29 & $\alpha$Sco   & i & 2006-269T07:41 &   -166$^\circ$ & 712 & 0.10 & +0.24 & \\
30 & RLeo                & i & 2006-285T02:11 &  -39$^\circ$  & 75 & 0.44 & +0.59 & Xd,C,I\\
30 & RLeo                & e & 2006-285T02:39 &  -85$^\circ$ &  74 & 0.44 & -0.08 & Xd,C,I\\
31 & CWLeo             & i & 2006-301T01:26 &  -23$^\circ$ &  167 & 3.07 & +1.21 & Xr,B\\
31 & CWLeo             & e &2006-301T02:10 & -103$^\circ$  & 188 & 3.07 & -0.91 & Xr,B\\
34 & $\alpha$Aur    & i & 2006-336T12:54 & 27$^\circ$ & 414 & 0.92 & +0.21 & F\\
36 & RHya                & i & 2007-001T17:07 & -165$^\circ$  & 333 & 0.35 & +0.43 & B\\
41 & $\alpha$Aur    & i & 2007-082T17:30 & 8$^\circ$ & 212 & 0.31 & -0.45 & B,I\\
41 & RHya                & i & 2007-088T07:24 & -156$^\circ$  & 112 & 0.09 & +0.18 & B,D,I\\
\hline
63 & RLeo                & i & 2008-094T13:11 &  96$^\circ$ & 368 & 0.23 & +0.93 & \\
63 & RLeo                & e & 2008-094T13:58 & 126$^\circ$ &  371 & 0.23 & +0.01 & G \\
65 & RCas                & i & 2008-112T00:43 & 30$^\circ$ & 77 & 0.30 & +0.27 & Xd,C,I \\
70 & CWLeo             & i & 2008-155T14:22 & 82$^\circ$ &  395 & 0.76 & -2.11 & \\
70 & CWLeo             & e & 2008-155T16:15 & 137$^\circ$ & 363 & 0.75 & -4.15 & \\
71 & $\gamma$Cru & i & 2008-160T09:04 & -173$^\circ$ & 422 & 0.28 & +0.46 & \\
72 & $\gamma$Cru & i & 2008-167T12:21 & -174$^\circ$ & 639 & 0.28 & +0.83 & \\
73 & $\gamma$Cru & i & 2008-174T15:30 & -174$^\circ$ & 635 & 0.28 &  +0.46 & \\
75 & RLeo                 & i & 2008-191T04:43 & 80$^\circ$ & 279 &  0.36 & +1.07 &\\
75 & RLeo                 & e & 2008-191T06:27 &138$^\circ$  &  279 & 0.36 & -0.63 & \\
77 & RLeo                 & i & 2008-205T06:58 &84$^\circ$  & 295 & 0.31 & +0.10 & \\
77 & RLeo                 & e & 2008-205T08:27 & 133$^\circ$  &  296 & 0.31 & +0.18 & \\
78 & $\gamma$Cru & i & 2008-209T20:07 & -175$^\circ$ & 304 & 0.14 & +0.52 & B\\
78 & $\beta$Gru       & i & 2008-210T09:35 & -96$^\circ$ &  254 & 0.34 & -0.26 & \\
80 & RSCnc              & i & 2008-226T02:03 & 57$^\circ$ & 310 & 0.56 & -0.61 & H\\
80 & RSCnc              & e & 2008-226T07:29 &155$^\circ$ & 305 & 0.56 & +0.00 & \\
81 & $\gamma$Cru & i & 2008-231T06:57 &-177$^\circ$ & 579 & 0.27 & +0.12 & \\
82 & $\gamma$Cru & 1 & 2008-238T15:32 & -177$^\circ$ &  721 & 0.27 & -0.04 & \\
85 & RSCnc              & i & 2008-262T22:29 & 59$^\circ$ & 306 & 0.54 & -0.30 & \\
85 & RSCnc              & e & 2008-263T03:48 &153$^\circ$ & 312 & 0.54 & +0.31 & \\
86 & $\gamma$Cru & i & 2008-268T03:12 &-178$^\circ$ & 1020 &0.41 & -0.33 & \\
87 & RSCnc              & i & 2008-277T16:16 & 59$^\circ$ & 326 & 0.53 & -0.12 & \\
87 & RSCnc              & e & 2008-277T21:30 &152$^\circ$ & 325 &0.53 & +0.01 & \\
89 & $\gamma$Cru & i & 2008-290T04:23 &-178$^\circ$ & 704 & 0.27 & -0.13 & B \\
92 & RSCnc              & i & 2008-315T01:56 &85$^\circ$ & 225 & 0.21 & +0.67 & B\\
93 & $\gamma$Cru & i & 2008-320T16:31 & -157$^\circ$ & 484 & 0.18 & +0.09 & B,E\\
94 & $\gamma$Cru & i & 2008-328T01:23 &-168$^\circ$ & 267 &0.13 & +0.30 & B,E\\
94 & $\epsilon$Mus & i & 2008-328T08:20 & -102$^\circ$ &  188 & 0.19 & +0.53 & B \\
94 & $\epsilon$Mus & e & 2008-328T13:02 & -44$^\circ$ & 188 & 0.19 & +0.02 & B\\
96 & $\gamma$Cru & i & 2008-343T11:45 & -172$^\circ$ & 236 & 0.14 & +0.18 & B\\
97 & $\gamma$Cru & i & 2008-351T11:03 & -172$^\circ$ & 794 & 0.42 & +0.44 & \\
\hline
\end{tabular}}}

\medskip
\small
Note times, longitudes and maximum DN are evaluated inside the Cassini Division.

Xr: Excluded from Cassini Division analysis due to coarse radial sampling

Xd: Excluded from Cassini Division analysis due to low signal levels

A: Low inclination occs, Saturn pole position adjusted to achieve approximate match with circular features

B: Data smoothed by 3 prior to edge-finding

C: Data resampled and smoothed by 5 prior to edge-finding

D: Data smoothed by 10 prior to finding Bessel Gap Inner Edge

E: Data smoothed by 12 prior to finding Laplace Gap Inner Edge

F: Jeffreys Gap Inner edge falls in data gap

G: Laplace Ringlet Inner edge falls in data gap

H: Kuiper Gap Inner edge falls in data gap

I: Spectral channels covering the range from 2.73-3.11 microns used.
\normalsize
\end{table}

Up through the end of 2008, VIMS has observed over 50 occultations, which have
yielded a total of 48 potentially useful cuts through the Cassini Division.
Table~\ref{obstab} lists the occultation cuts used in this analysis,
along with the occultation times and inertial longitudes of the
observations, the maximum Data Number (DN) detected, the radial
sampling scale and the shift required to bring circular features
into alignment (see below).

\subsection{Geometrical Navigation}

\begin{table}[tbp]
\caption{Assumed star positions in the International
Celestial Reference Frame at the Hipparcos epoch. Data obtained from the
Simbad astronomical database and the Hippacos online
catalog. Note that all the stars except CWLeo and 
$\eta$Car were present in the Hipparcos catalog.}
\label{starpos}
\begin{tabular}{|l|c|c|c|c|c|}\hline
Star & Right Ascension & Declination & Proper Motion & Proper Motion & Parallax$^a$ \\
& hr:min:sec & hr:min:sec & R.A. (mas/yr) & Dec. (mas/yr) & mas \\ \hline
$o$Cet & 		02:19:20.7866 &	-02:58:37.418 & 	+10.33 & -239.48 & 7.79 \\
$\alpha$Sco & 16:29:24.4675 &	-26:25:55.006 & 	-10.16 & -23.21 & 5.40 \\
$\alpha$Ori &	05:55:10.2892 &	+7:24:25.331 &		+27.33 & +10.86 & 7.63 \\
$\alpha$Tau &	04:35:55.2005 &	+16:30:35.142& 	+62.78 & -189.36 & 50.09 \\
$\delta$Vir & 	12:55:36.4833 & 	+03:23:51.355 &	-471.44 & -52.81 & 16.11 \\
RLeo	& 	09:47:33.4907 &	+11:25:44.020 &	-0.57	 & -42.70	& 9.87 \\
CWLeo 	& 	09:47:57.382 &		+13:16:43.66 &		-- & -- & -- \\		
$\alpha$Aur & 	05:16:41.2956 &	+45:59:56.505 &	-75.52 & -427.13 & 77.29 \\
RHya	& 	13:29:42.8189 &	-23:16:52.888 &	-60.73 & +11.01 & 1.62 \\
RCas	&	23:58:24.7936	&	+51:23:19.545 &	+84.39 & +18.07 & 9.37 \\
$\gamma$Cru&12:31:09.9293 &	-57:06:45.249 & 	+27.94 & -264.33 & 37.09 \\
$\eta$Car &	10:45:03.591 &		-59:41:04.26 &		-7.6 & +1.0 & -- \\			
$\beta$Gru &	22:42:39.9349 &	-46:53:04.437 &	+135.68 & -4.51 & 19.17 \\
RSCnc &		9:10:38.8054 &		+30:57:47.589 &	-9.41	 & -33.05 & 8.21 \\
$\epsilon$Mus &12:17:34.6363 &	-67:57:38.418 &	-231.26 & -26.37 & 10.80 \\ \hline
\end{tabular}

$^a$ Parallax measured at Earth
\end{table}

Table~\ref{starpos} gives the assumed positions of the stars used in this 
analysis in the International Celestial Reference Frame. For each occultation, the position 
of the star is adjusted to account for both the proper motion of the star 
and the parallax at Saturn. The available SPICE kernels provided by the Cassini navigation team 
were  then used to predict the apparent position (radius and inertial
longitude) of the star in Saturn's ring plane as a
function of time in a planetocentric reference frame, taking into account stellar aberration. In nearly all cases, this estimate of the occultation geometry was
confirmed to be accurate to within a few kilometers using the known
radii of nearly circular gap edges in the Cassini Division and the outer A Ring
(Features 1, 3, 4, 13, 16 and 20 of French {\em et al.} 1993). The exceptions
were the low-inclination stars $o$ Ceti and $\delta$ Virginis, for which
features could be tens of kilometers away from their nominal positions.
In these cases, the fiducial position of Saturn's pole was adjusted slightly 
(by at most 0.015$^\circ$) to bring these cuts into alignment with the other
occultations. The residual scatter in the estimated 
locations of circular features is consistent with uncertainties in
the star positions, observation timing and spacecraft trajectory (see below).

\subsection{Lightcurve Generation} 

While each occultation produces several time series of brightness
measurements at multiple wavelengths, only the average brightness
around a wavelength of three microns is used for this analysis.
This is because in general, the signal measured by the  
VIMS instrument was a  combination of transmitted starlight 
and reflected sunlight. The reflected signal from the rings is strongly 
attenuated at wavelengths close to the strong 2.9 micron 
water-ice absorption band, so using the spectral channels
in this range minimizes the ring background. 

To save on data volume, in many of these occultations the normal spectral 
resolution of the instrument was reduced by a factor of eight. A 
total of 32 spectral  channels were returned with an average resolution 
$\Delta\lambda \simeq 0.13 \mu$m. For the few occultations taken at 
full spectral resolution, the
data from the appropriate spectral channels were co-added in software 
after the fact to make their spectral resolution and signal-to-noise consistent
with the other occulations.  For most of the occultations,  the summed
spectral channel corresponding to the wavelength range between 
2.87 and 2.98 microns was used in this analysis. However, a small 
number of occultations (Rev 13 $\alpha$ Scorpii, 
Rev 30 R Leonis, Rev 41 $\alpha$ Auriga, Rev 41 R Hydra 
and Rev 65 R Cassiopaea), had either sufficiently low signal-to-noise ratios or
sufficiently high cosmic-ray-fluxes that a single profile was too noisy to
obtain reliable edge detections. In these cases, three brightness profiles
were averaged together, covering the wavelength range of 2.73-3.11 
microns. This simple averaging reduced the instrumental noise
and cosmic ray background to acceptable levels.

While using data near 2.9 microns minimizes the background
from the illuminated rings, it does not completely eliminate
this signal from all occultations. This residual ring background
can be eliminated by comparing data at different wavelengths
\citep{NH09}, but such refinements are not necessary 
here because the goal of this analysis is simply to determine the locations of 
sharp edges. The time variations in the
ring background are limited by the {\it spatial} resolution of the
instrument, while the time variations in the stellar signal are
limited by the {\it temporal} resolution, so the ring background varies
slowly with time compared to the variations in the stellar signal
and has negligible effect on the inferred locations of sharp edges.

\subsection{Edge Detection}

After deriving the lightcurve and geometrically navigating each 
occultation, we determined the radius, inertial longitude and time 
when the star passed behind 22 edges in the Cassini Division:
the inner and outer edges of the Huygens, Herschel, Russell, Jeffreys,
Kuiper, Laplace, Bessel and Barnard Gaps, as well as the inner and outer
edges of the dense ringlets in the Huygens, Herschel and Laplace Gaps.
Note that we do not consider the ``Strange" ringlet in the Huygens Gap in 
this analysis, nor the low-optical depth ringlets in the Huygens, Jeffreys
or Laplace Gaps.

The edges in the Cassini Division are quite variable in their
morphology, and even a single edge may have different shapes in 
different occultations. This could make the identification of edges
based on half-light levels \citep{French93} problematic, so instead
we identified the edge as the point in the lightcurve with the
steepest slope. Even so, some effort was required to prevent the edge-detection
algorithm from mistaking cosmic ray spikes or other sharp ring features
for the desired edge.  The algorithm therefore begins by normalizing the
total signal (star + reflected ringshine, if any) to unity in the gaps. This 
is accomplished by first determining the
average DN in two clear zones (117,700-117,750 km and 
119,970-120,000 km) within the two widest gaps in the Cassini Division:  
the Huygens Gap and the Laplace Gap. A linear trend based on these two
numbers establishes the unocculted stellar brightness throughout this region
and is used to normalize the data so that the signal in each gap is approximately
1.0. 

For each edge, a fiducial zone is selected for analysis which is sufficiently
wide to accommodate all observed edge positions and residual pointing errors.
(For the B ring edge this region is 150 km wide. For the Huygens ringlet edges
this zone is 40 km wide. For the Herschel Ringlet and Herschel Gap inner edge it is
20 km wide. For the inner edge of the Laplace gap it is 15 km wide. For the 
Herschel gap outer edge, the Russell and Jeffreys gap inner edges, the 
Laplace ringlet and the Laplace gap outer edge, and the Bessel and 
Barnard gaps it is 10 km. For the Russell and Jeffreys gaps' outer edges
and the Kuiper gap edges it is 5 km.)  The algorithm first makes a preliminary estimate
of the location of the relevant gap edge based on where the signal in this
region first deviates significantly from unity. The final estimate of the edge location
is then the point where the brightness profile has the steepest slope in a 
region within $\pm$4 km of the preliminary edge estimate
(except for the low-resolution $\delta$ Virginis occultation, where a
region $\pm$20 km was used). 

For certain occultations with fast time-sampling or low signal-to-noise, the 
raw profile is too noisy for the above algorithm to find all the edges reliably. 
In these cases, a boxcar smoothing was applied to the data prior to estimating
the edge position. The occultations where this was done and the specific
smoothing lengths used are given in Table~\ref{obstab}. In all cases, the 
uncertainty in the edge position is determined by the radial sampling interval
because this is typically much larger than the projected stellar diameter or the fresnel zone.

\subsection{Occultation Quality}

While the above procedure was applied to all the occultation cuts in Table~\ref{obstab},
there were a few occultations which we elected not to use with the 
Cassini Division edges. The 
Rev 30 R Leonis and  Rev 65 R Cassiopaea occultations had 
signal-to-noise ratios too low to detect all the edges in the Cassini Division reliably, while
the resolution of the Rev 29 $\delta$ Virginis and the Rev 31 CW Leonis occultations were so much lower than 
the rest of the observations that they would not contribute much to our understanding of
most of these edges. These data are therefore not considered ``Quality" occultations
and are not used in the analysis involving the edges within the Cassini Division 
itself (they are designated with Xr or Xd in Table~\ref{obstab}). However, many of these occultations occur at early times in the Cassini mission, 
when VIMS occultations are comparatively rare, so they are useful when considering 
the time-evolution of the B ring edge (see below).

\section{Data Analysis}

\begin{table}
\caption{Elementary Properties of Cassini Division Gap Edges}
\label{gapov}
\begin{tabular}{|l c c l|}
\hline
Feature & Mean Radius & St.Dev. & Edge Type \\
& (km) & (km) & \\
\hline
B-ring Outer Edge                   & 117564.4	& 51.4 & Resonant (2:1 Mimas) \\
Huygens Ringlet Inner Edge & 117804.4 & 20.3 & Eccentric \\
Huygens Ringlet Outer Edge & 117822.8  & 20.2 & Eccentric \\
Huygens Gap Outer Edge      & 117930.6 & 2.8 & Unknown \\
\hline
Herschel Gap Inner Edge      & 118188.2 & 5.9 & Eccentric \\
Herschel Ringlet Inner Edge & 118233.9 & 2.3 & Unknown \\
Herschel Ringlet Outer Edge & 118263.2 & 2.7 & Unknown \\
Herschel Gap Outer Edge      & 118283.4 & 1.1 & Unknown (Circular?) \\
\hline
Russell Gap Inner Edge          & 118589.7 & 5.2 & Eccentric \\
Russell Gap Outer Edge         & 118628.2 & 0.9 & Circular \\
\hline
Jeffreys Gap Inner Edge         & 118929.6 & 2.7 & Eccentric \\
Jeffreys Gap Outer Edge        & 118966.5 & 0.8 & Circular \\
\hline
Kuiper Gap Inner Edge          & 119401.7 & 1.0 & Eccentric \\
Kuiper Gap Outer Edge         & 119406.1 & 0.7 & Circular \\
\hline
Laplace Gap Inner Edge      & 119845.2 & 2.6 & Eccentric \\
Laplace Ringlet Inner Edge & 120036.4 & 1.8 & Unknown (Resonant? 9:7 Pandora) \\
Laplace Ringlet Outer Edge & 120077.7 & 2.1 & Eccentric \\
Laplace Gap Outer Edge      & 120085.7 & 1.3 & Eccentric \\
\hline
Bessel Gap Inner Edge         & 120231.3 & 1.6 & Eccentric \\
Bessel Gap Outer Edge        & 120243.6 & 1.0 & Circular? \\
\hline
Barnard Gap Inner Edge       & 120303.7 & 2.4 & Resonant? (5:4 Prometheus) \\
Barnard Gap Outer Edge      & 120315.9 & 0.7 & Circular \\
\hline
\end{tabular}
\end{table}

The above procedures yield roughly 40 ``quality" estimates of the position of each
of the 22 selected gap edges for a range of different times and 
inertial longitudes. The simplest statistics that can be computed from these data
are the (unweighted) means and standard deviations of all the edge estimates, which
are presented in Table~\ref{gapov}. Of particular interest are the standard deviations, 
which range from around 1 km up to 50 km for the edge of the B ring. 

Based on these variances, we can begin to identify several different classes
of edges in the Cassini Division. First of all, there are several features whose
scatter is around 1 km, which likely represent truly circular ring edges. These 
include the outer edges of the Russell, Jeffreys, Kuiper, Barnard and possibly
the Bessel and Herschel Gaps. (The inner edge of the Kuiper gap has a similarly
low variance, but further investigation shows that it belongs in the eccentric class
of ring features, discussed below). 

The remaining edges all have noticeably larger variances and therefore appear
to be non-circular in some form. Three of these edges (the outer edge of the B ring, and the
inner edges of the Laplace Ringlet and the Barnard Gap) are close to mean-motion 
resonances with known satellites of Saturn (Mimas 2:1, Pandora 9:7
and Prometheus 5:4, respectively) which could provide natural explanations for 
their shapes. However, this still leaves most of the edges unexplained. We have found that
10 of these features (both edges of the Huygens ringlet, the inner edges
of the Herschel, Russell, Jeffreys, Kuiper, Laplace and Bessel Gaps, and the 
outer edges of the Laplace gap and ringlet) can be well fit with ellipses that
precess around the planet at the expected rate given Saturn's oblateness. Four non-circular
 edges (both sides of the Herschel ringlet, as well as the outer edges of the 
 Huygens and possibly the Herschel Gaps) cannot be fit by simple elliptical
 models and therefore require further investigation. 

We will consider each of these different classes in detail below.
First, we will investigate the apparently circular edges and show how they can be used
to refine the estimates of the other edge locations. Next, we will discuss the eccentric 
features and compare the amplitudes and precession rates of these features. 
We will then consider the inner edge of the Barnard Gap and explore whether
the structure of this edge can be explained by the perturbations from  
Prometheus. Finally, we will look at the
outer edge of the B ring, which has the largest radial excursions and the most
complex structure. The inner edge of the Laplace ringlet, both edges of the Herschel
ringlet, and the outer edges of the Herschel and Huygens Gaps still require further
study and will not be discussed in detail in this paper.

\subsection{Circular Edges} 

\begin{figure}
\centerline{\resizebox{5in}{!}{\includegraphics{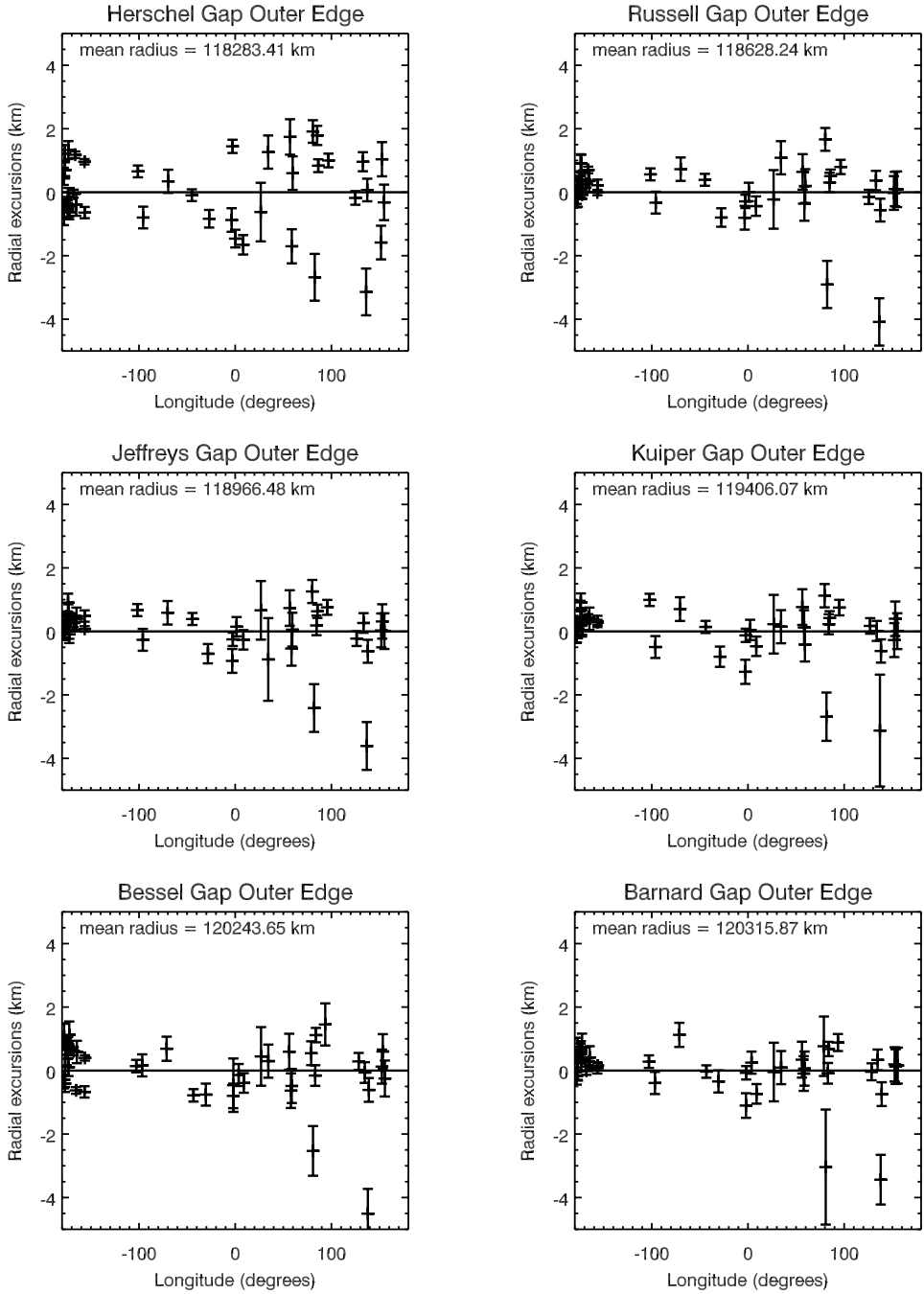}}}
\caption{The radial excursions (deviations from the unweighted mean) of six apparently
circular edges in the Cassini Division, plotted as functions of inertial longitude. The error bars 
indicate the radial sampling scale of each occultation.}
\label{cdcirc}
\end{figure}

Figure~\ref{cdcirc} shows the scatter in the inferred positions of the outer edges of the Herschel,
Russell, Jeffreys, Kuiper, Bessel and Barnard Gaps. All of these data sets have
rather low dispersions. Furthermore, there is a strong correlation among the radial excursions of different edges from the same occultation. (For example, the low points in all six panels
near $80^\circ$ and  140$^\circ$ longitude come from the Rev 70 CW Leo occultations. 
Since this object was not observed by Hipparcos, the occultation geometry 
here is more uncertain than the others.) This suggests that most of the scatter in 
the positions of these edges is due to small errors in the geometric reconstructions
of different occultations.

Assuming that most of these edges are circular, we use these features to
compute a radial offset for each cut and refine the location estimates for the 
remaining edges. For each occultation, we compute the mean deviation 
of the measured positions of the outer edges of the Russell, 
Jeffreys, Kuiper, Bessel and Barnard gaps from the mean values
listed in Table~\ref{gapov} (118628.2, 118966.5, 119406.1, 120243.6 and 
120315.9 km, respectively). These radial shifts, tabulated in Table~\ref{obstab},
are then applied to all edges for that occultation. Note that the 
nominal positions of the outer edges of the Russell, Jeffreys and Bessel
gaps are 1-2 km interior to the values reported in French et al. (1993).
Such a small shift should be of no major consequence for this analysis, 
but will be the subject of a future investigation of Saturn's pole orientation 
 incorporating data from all circular features in the rings.

\subsection{Eccentric Edges}

The observed radial position $r'$ of an eccentric edge depends on the inertial 
longitude $\lambda$ and time relative to some epoch $\delta t$ 
of the observation as follows:
\begin{equation}
r'=r'_o-A*\cos(\lambda-\Omega_p\delta t-\lambda_o),
\label{ecceq}
\end{equation}
where $r'_o$ is the mean edge location, $A$ is the amplitude of the radial excursions,
$\lambda_o$ is a phase offset and $\Omega_p$ is a pattern speed. Since an eccentric edge
closely follows the path of a freely precessing particle on an eccentric orbit, we expect
$\Omega_p = \dot{\varpi}$, where $\dot{\varpi}$ is the apsidal precession rate, which in the 
Cassini Division ranges between 4.5$^\circ$/day and 5.1$^\circ$/day. 

Preliminary investigations showed that ten of the edges in the Cassini Division could
be well fit by such an eccentric model with $\Omega_p$ close to the expected 
value of $\dot{\varpi}$. These features are both edges of the Huygens ringlet, the
 inner edges of the Herschel,
Russell, Jeffreys, Kuiper, Laplace and Bessel Gaps, as well as the outer edges of the Laplace
Gap and ringlet. The Huygens ringlet was previously known to be an eccentric structure
\citep{Turtle91, Spitale06}, and there was some evidence that the inner
edge of the Herschel Gap might be as well \citep{FC89}. However, the fact that a majority 
of the edges in the Cassini Division are simple ellipses is a rather unexpected finding.

Having identified these features, we sought to determine the parameters
in Equation~\ref{ecceq}. Given the extensive occultation data available, and mindful
that the precise precession rates at a given location might be affected by
nearby ring material \citep{NP88}, we chose not to assume a predicted
pattern speed for each edge but instead included $\Omega_p$ as a free parameter
in each fit.

For each possible value of the pattern speed ${\Omega}_p$, we computed the following quantities:
\begin{equation}
\alpha_R=\frac{1}{n}\sum_{i=1}^n(r'_i-\bar{r}')*\cos(\lambda_i-\Omega_p\delta t_i)
\end{equation}
\begin{equation}
\alpha_I=\frac{1}{n}\sum_{i=1}^n(r'_i-\bar{r}')*\sin(\lambda_i-\Omega_p\delta t_i)
\end{equation}
where $\bar{r}'$ is the mean radius of the edge and the sums
are over the $\sim$40 measurements of $r'$. If the radial position of
the edge is described by Equation~\ref{ecceq}, then $\sqrt{\alpha_R^2+\alpha_I^2}=\rho A/2$, 
where $\rho$  is the correlation coefficient between $r'_i-\bar{r}'$ and  
$\cos(\lambda_i-\Omega_p\delta t_i-\lambda_o)$. This function
will be at a maximum when there is perfect correlation between
these parameters (i.e. when $\rho=1$). In the limit of perfect sampling of
all possible true 
anomalies ($\lambda_i-\Omega_p\delta t_i$), this  should only 
occur when the assumed pattern speed $\Omega_p$
equals the true pattern speed. Thus the value of $\Omega_p$
that yields the highest value of $\sqrt{\alpha_R^2+\alpha_I^2}$ provides the
best estimate of the true pattern speed. In this case, the parameters 
$\alpha_R$ and $\alpha_I$ will approach $-(A/2)\cos\lambda_o$ and $-(A/2)\sin\lambda_o$ 
(again in the limit of perfect sampling of all possible true anomalies), 
so we can estimate the amplitude and phase parameters using 
the equations $A/2=\sqrt{\alpha_R^2+\alpha_I^2}$ and 
$\tan \lambda_o=\alpha_I/\alpha_R$.

\begin{figure}[tbp]
\centerline{\resizebox{5in}{!}{\includegraphics{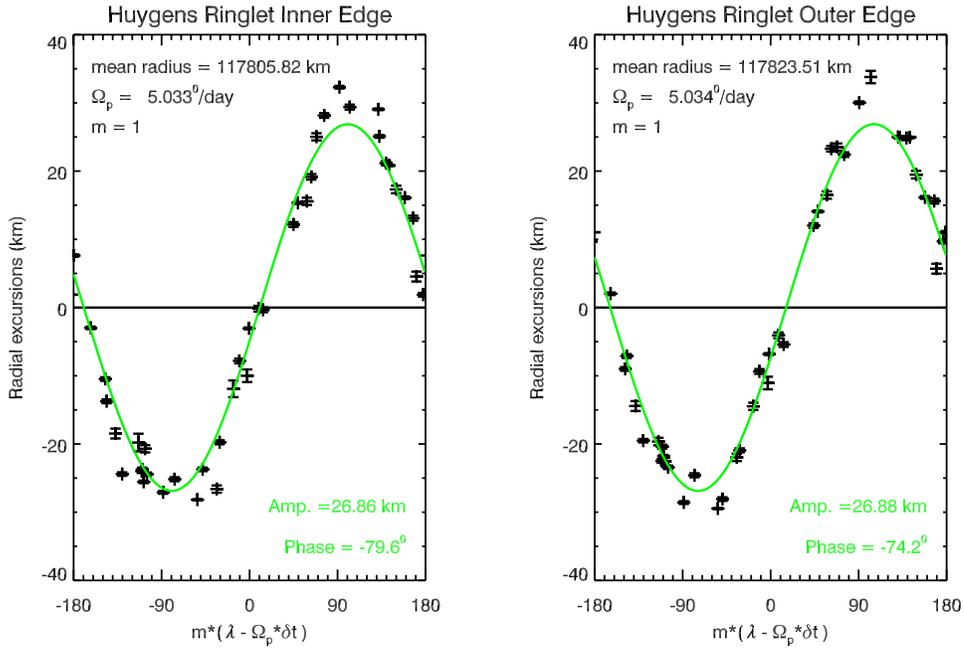}}}
\caption{Radial excursions (after offset subtraction) of the inner and outer edges of the Huygens
Ringlet. Error bars indicate the radial sampling scales of the different occultations.
Note time is measured relative to an epoch of 2005-195T02:13:13.557 (which corresponds to a Cassini spacecraft clock time of 150000000), and the mean radii are for the eccentric model rather than the data.}
\label{m1huygens}
\end{figure}

\begin{figure}[htbp]
\centerline{\resizebox{5in}{!}{\includegraphics{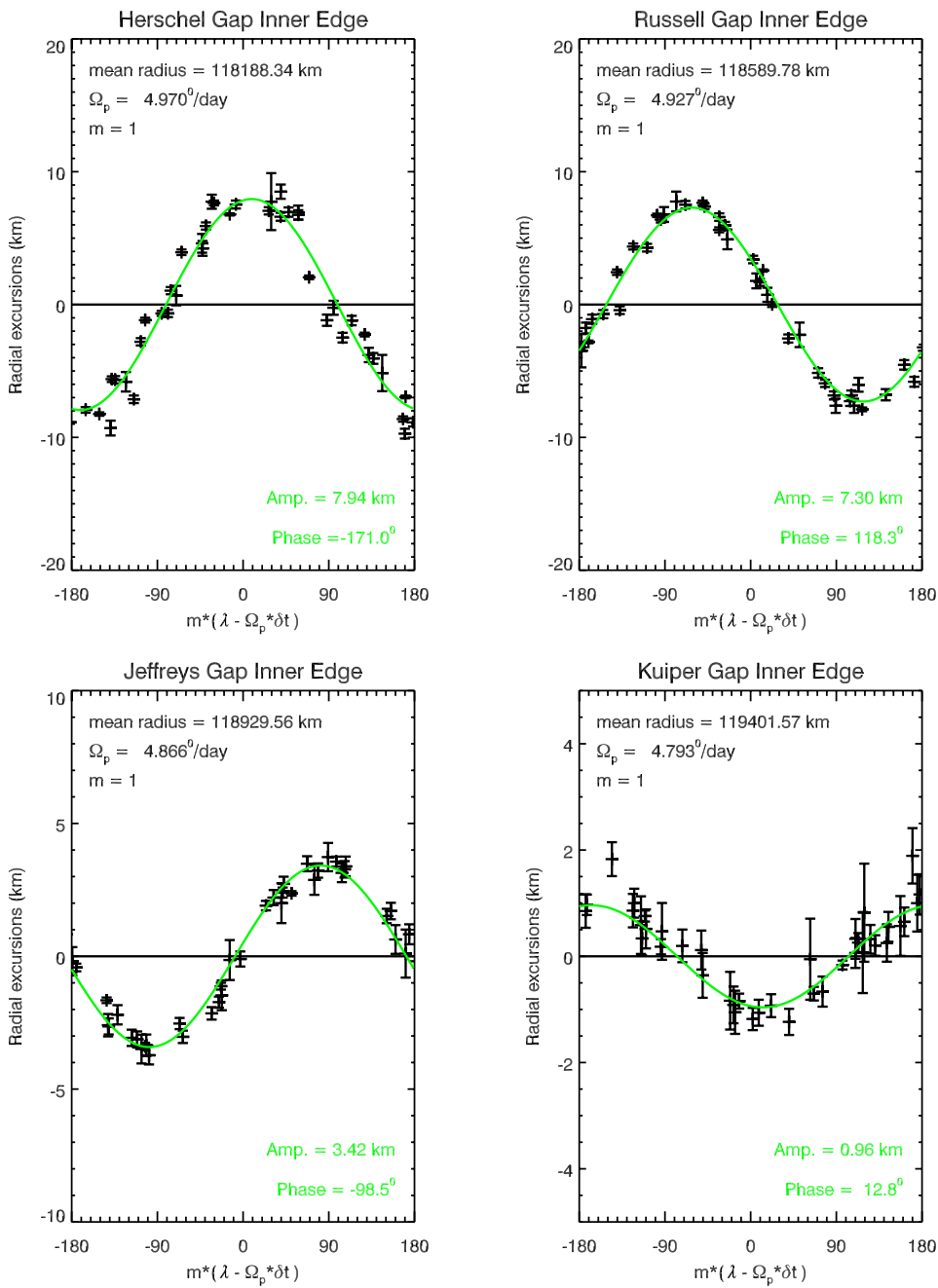}}}
\caption{Radial excursions (after offset subtraction) of the inner edges of the Herschel, Russell,
Jeffreys and Kuiper gaps. Error bars indicate the radial sampling scales 
of the different occultations. Note time is measured relative to an epoch of 2005-195T02:12:13.557 (which corresponds to a Cassini spacecraft clock time of 150000000), and the mean radii are for the eccentric model rather than the data.}
\label{m1g4g7}
\end{figure}

\begin{figure}[htbp]
\centerline{\resizebox{5in}{!}{\includegraphics{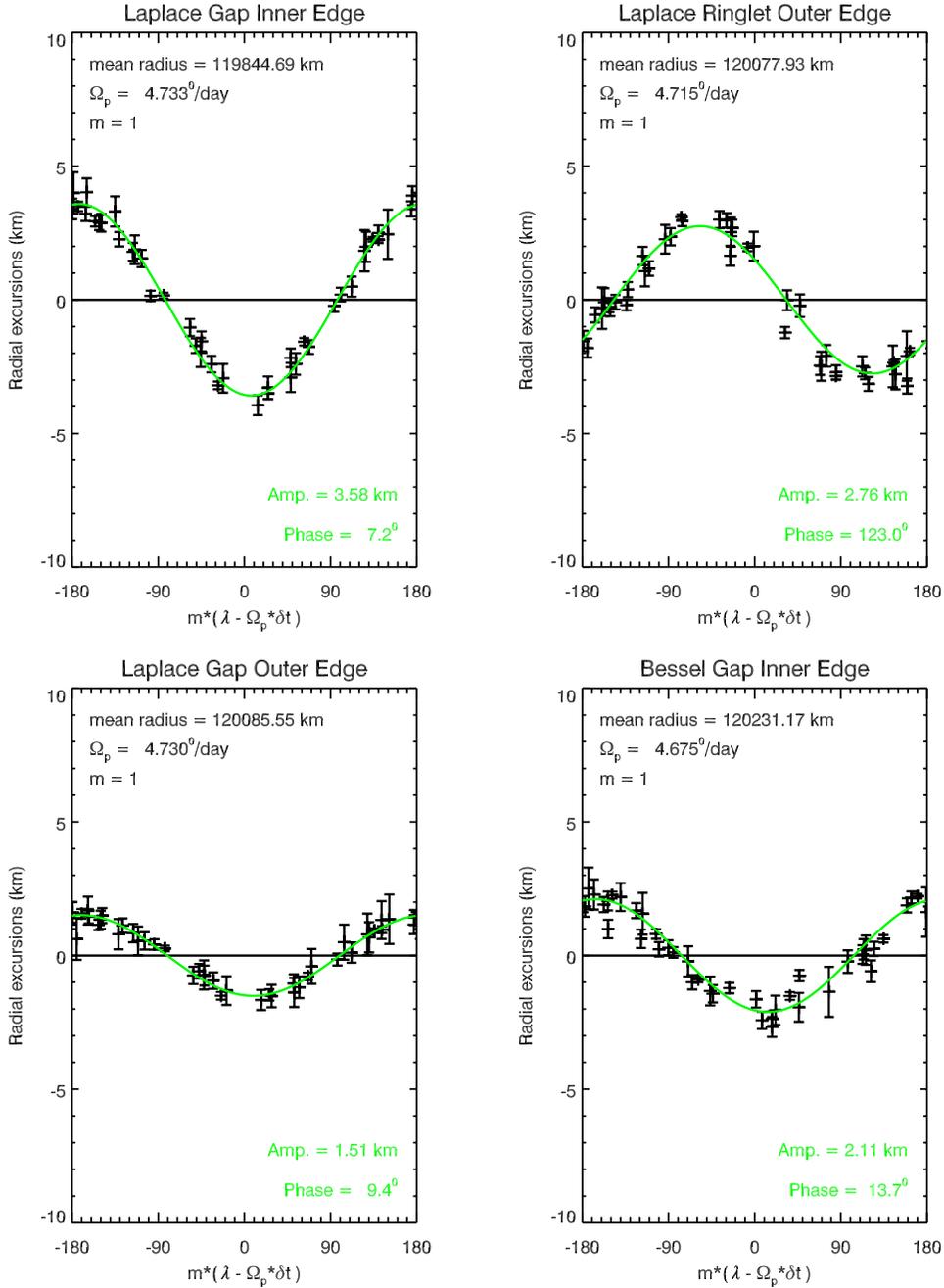}}}
\caption{Radial excursions (after offset subtraction) of the inner edges of the Laplace and Bessel Gaps, 
as well as the outer edges of the  of the Laplace Gap and Ringlet. Error bars indicate 
the radial sampling scales of the different occultations. Note time is measured relative to an epoch of 2005-195T02:12:13.557 (which corresponds to a Cassini spacecraft clock time of 150000000),
and the mean radii are for the eccentric model rather than the data.}
\label{m1g8g10}
\end{figure}

Figures~\ref{m1huygens}-~\ref{m1g8g10} show the measured radial excursions
for the eccentric edges as a function of $\lambda-\Omega_p\delta t$ using the best-fit 
pattern speeds. In all cases, the data are well organized into a sine wave. Table~\ref{ecctab}
lists the best fit parameters for these edges. Note that the observed pattern speeds
of these features are not far from the predicted pattern speeds derived by assuming
the edge follows a freely precessing elliptical orbit. The uncertainty in the observed 
pattern speeds is set by the time baseline of the observations, which is approximately
1300 days. A difference of 0.01$^\circ$/day in the pattern speed would therefore
shift the phase of the first and last occultations by roughly 15$^\circ$, which is
probably on the edge of detectability. 

\begin{table}[tbp]
\caption{Parameters for the Eccentric Edges}
\label{ecctab}
\resizebox{6in}{!}{\begin{tabular}{|c c c c c c|}
\hline
Feature & Mean Radius$^a$ & Amplitude & Phase$^b$ & Pattern Speed & Pattern Speed \\
& & & & (Observed) & (Predicted)$^c$ \\ \hline
Huygens Ringlet Inner Edge & 117805.8 km & 26.9 km & -80$^\circ$ & 5.03$^\circ$/day 
& 5.022$^\circ$/day \\
Huygens Ringlet Outer Edge & 117823.5 km & 26.9 km & -74$^\circ$ & 5.03$^\circ$ /day 
& 5.019$^\circ$/day \\
\hline
Herschel Gap Inner Edge & 118188.3 km & 7.9 km & -171$^\circ$ & 4.97$^\circ$/day 
& 4.964$^\circ$/day\\
\hline
Russell Gap Inner Edge & 118589.8 km & 7.3 km & 118$^\circ$ & 4.93$^\circ$/day 
& 4.904$^\circ$/day\\
\hline
Jeffreys Gap Inner Edge & 118929.6 km & 3.4 km & -98$^\circ$ & 4.87$^\circ$/day 
& 4.854$^\circ$/day\\
\hline
Kuiper Gap Inner Edge & 119401.6 km & 1.0 km & 13$^\circ$ & 4.79$^\circ$/day 
&4.786$^\circ$/day\\
\hline
Laplace Gap Inner Edge & 119844.7 km & 3.6 km & 7$^\circ$ & 4.73$^\circ$/day 
&4.723$^\circ$/day\\
Laplace Ringlet Outer Edge & 120077.9 km & 2.8 km & 123$^\circ$ & 4.72$^\circ$/day 
& 4.691$^\circ$/day\\
Laplace Gap Outer Edge & 120085.6 km & 1.5 km & 9$^\circ$ & 4.73$^\circ$/day 
& 4.689$^\circ$/day\\
\hline
Bessel Gap Inner Edge & 120231.2 km & 2.1 km & 14$^\circ$ & 4.68$^\circ$/day 
& 4.669$^\circ$/day\\
\hline 
\end{tabular}}

\medskip

$^a$ Note these  mean radii differ slightly from those in Table~\ref{gapov} because 
these are the mean radii of the eccentric model rather than the mean of the data.

$^b$ Longitude of edge's pericenter for an eccentric model using the given
pattern speed at an epoch of 2005-195T02:12:13.557 
(which corresponds to a Cassini spacecraft time of 150000000).

$^c$ Predicted precession rate of eccentric particle orbit with semi-major axis
equal to the observed mean edge radius, based on current
estimates of Saturn's gravity field parameters 
\citep{Jacobson06}

\end{table}

What is particularly interesting about the fitted pattern speeds of the eccentric
edges is that they almost form a regular sequence. To see this, first note that both edges of the
Huygens Ringlet have the same pattern speed (5.03$^\circ/$day), as expected for a narrow 
ringlet, and that all the eccentric features in the Laplace Gap also have
similar pattern speeds (4.72$^\circ/$day), so there are only seven distinct
pattern speeds in Table~\ref{ecctab}, one for each gap/ringlet. The average
difference in pattern speed between adjacent features is 0.06$^\circ/$day, 
with a standard deviation of about $0.01^\circ/$day, which is comparable to the 
uncertainty in the precession rates of the patterns (outliers being the $0.04^\circ$/day
difference between the pattern speeds for the Herschel and Russell Gap inner edges 
and the $0.08^\circ$/day difference between the Jeffreys and Kuiper Gap inner edges). 
Also, with the exception of the almost-circular Kuiper Gap inner edge, there
is an almost monotonic decrease in the amplitude of these patterns from 27 km
at the Huygens Ringlet to 2 km at the Bessel Gap.
This regularity hints that these eccentric edges may be 
controlled by a series of closely spaced resonances. This idea will be explored
in more detail below after a consideration of the structure of two resonantly-controlled edges.

\subsection{Barnard Gap Inner Edge}

\begin{figure}[tbp]
\centerline{\resizebox{5in}{!}{\includegraphics{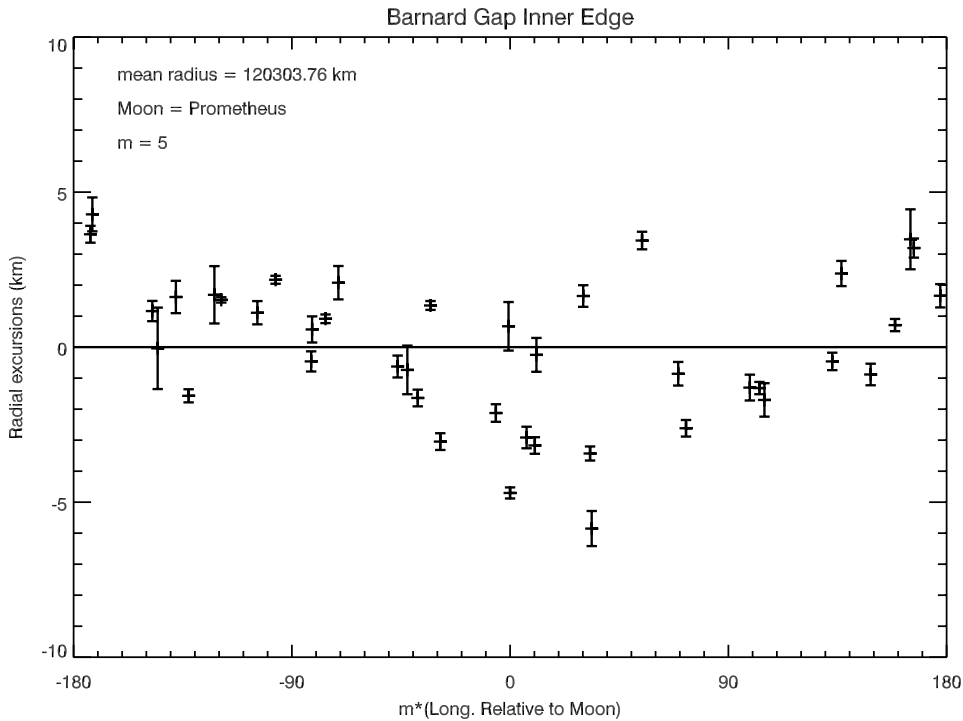}}}
\caption{Radial excursions of the inner edge of the Barnard Gap, plotted
as a function of five times the longitude relative to Prometheus. Note that the
data are reasonably well organized in this longitude system, consistent with
the edge being influenced by the 5:4 resonance with Prometheus.}
\label{promedge}
\end{figure}

The Barnard gap inner edge is a special case because it is the only
inner edge of a gap in the Cassini Division besides the B-ring edge that cannot be fit
to a simple eccentric model. All the other non-circular, non-eccentric edges are either on 
ringlets within the gaps (Herschel and Laplace) or at the outer edges of gaps containing
such ringlets (Huygens and Herschel). Furthermore, the mean radius of the Barnard Gap's
inner edge is 120303.7 km, which is very close to the predicted location of
the Prometheus 5:4 Inner Lindblad Resonance at 120304.1 km. Thus it is reasonable to
expect that the shape of this edge is described by the following function:
\begin{equation}
r=r_o-A*\cos(5(\lambda-\lambda_{Prometheus})).
\end{equation}
Figure~\ref{promedge} plots the radial excursions of this edge as a function of
$5(\lambda-\lambda_{Prometheus})$, and indeed shows that most of the data
can be roughly described by the above functional form. There are some clear deviations from the
expected pattern, most noticeably some points with large positive excursions
near $5(\lambda-\lambda_{Prometheus})=20^\circ$, and a possible phase
shift in the data relative to the model. Similar deviations are seen
in the B ring edge as well (see below), and may reflect complications in the dynamics
of resonantly controlled edges. In spite of this, it does appear that
the Barnard Gap is strongly influenced, and probably controlled by, 
the 5:4 Lindblad resonance with Prometheus.

\subsection{B-ring Outer Edge Observations}

\begin{figure}
\centerline{\resizebox{5in}{!}{\includegraphics{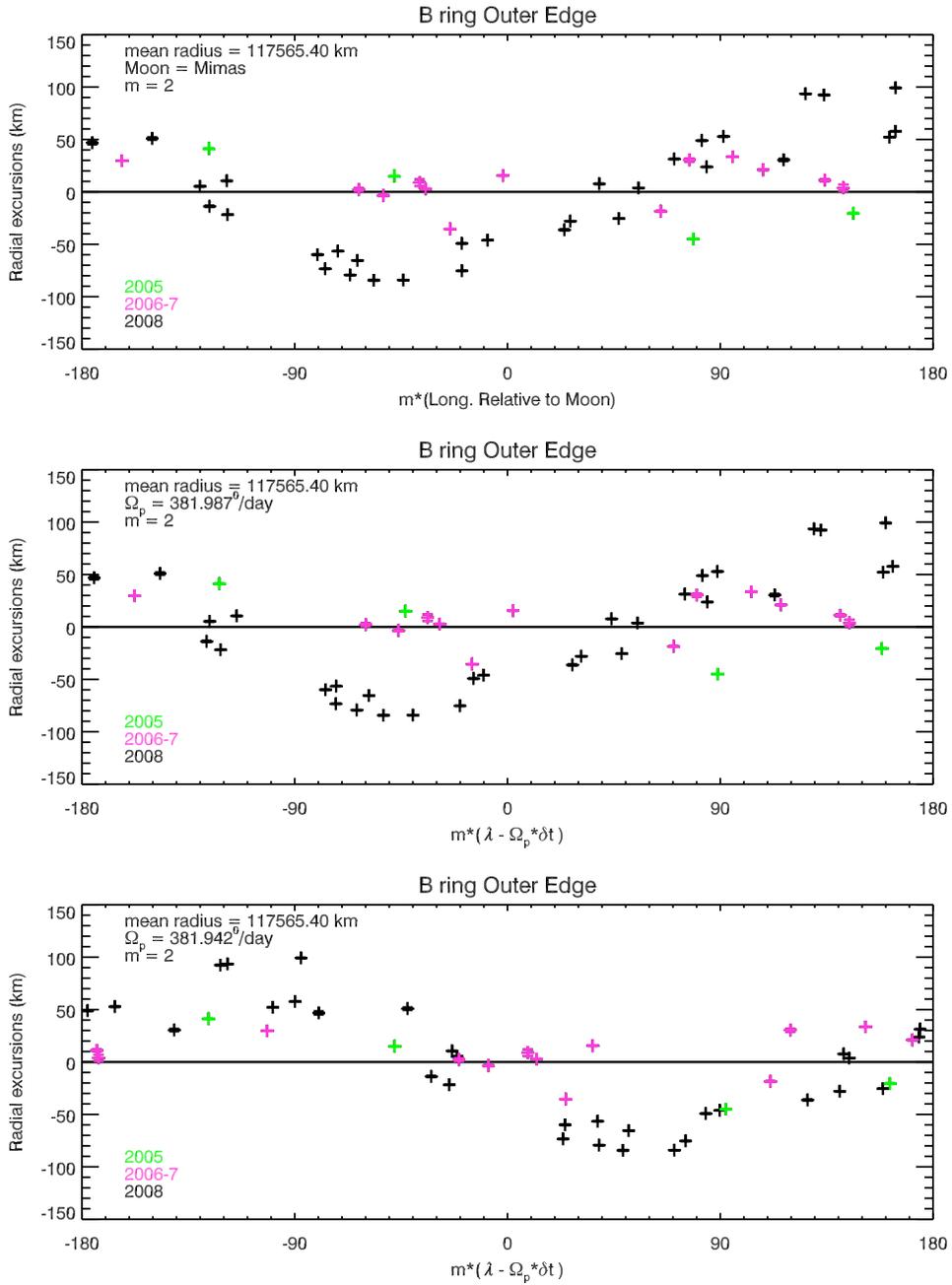}}}
\caption{Radial excursions of the outer edge of the B ring. The top panel
shows the data plotted versus the difference in longitude relative to Mimas.
The middle panel shows the data assuming a constant pattern speed
equal to Mimas' mean motion, and the bottom panel shows the data
organized using a slightly slower pattern speed that provides the best fit to the data. Note
that data from all 48 occultation cuts in Table~\ref{obstab} are used in these plots.}
\label{bedge2}
\end{figure}

The outer edge of the B ring has long been known to be strongly affected by the
Mimas 2:1 Inner Lindblad Resonance \citep{GT78, Smith82}. The radial excursions 
of a particle's orbit near this resonance can be described by the following equation
(see Section 4 below):

\begin{equation}
r=r_o-A*\cos(2(\lambda-\lambda_{Mimas}))
\end{equation}

Analyses of the Voyager data showed that the above expression provided
a reasonably accurate description of the outer edge of the B ring in 1981 
with an amplitude $A \simeq 75$ km \citep{Porco84}. However, as
shown in  the top panel of Figure~\ref{bedge2}, the VIMS occultation data do 
not fit this simple model so clearly. The measurements obtained in 2008 
do possess a clear  $m=2$  structure with an amplitude of 75 km, 
but there is a significant phase offset,  such that the minimum radius does not occur 
where $\lambda \simeq \lambda_{Mimas}$ but instead lags behind this point by about  
40$^\circ$ in phase (i.e. one of the two minima falls  $\sim$20$^\circ$ in longitude behind Mimas). 
Furthermore, the data taken prior to 2008 fall well away from  the curve 
described by the 2008 data. The radial differences between 
the 2005-2007 and 2008 data are as much as 100 km, which is far too large to be explained by
pointing errors, time variability in the pattern's amplitude or some additional (smaller
amplitude) perturbations  in the edge position.  Instead, the differences between the 
2005-2007 data and the 2008 data  are best explained by a time variation in the 
orientation of the pattern relative to Mimas. Indeed, if we assume the
pattern moves at a speed $\sim 0.045^\circ/$day slower than Mimas' mean 
motion, then the 2005-2007 data are well aligned with the 2008 data (see bottom panel
in Figure~\ref{bedge2}). It therefore appears that between 2005 and 2008  the phase
of this $m=2$-symmetric pattern has drifted backwards relative to Mimas.\footnote{
The long-term average mean motion of Mimas is 381.994509 $\pm 0.000005^\circ$/day 
\citep{HT93}, but the 4:2 resonance with Tethys results in a slow variation in the mean motion 
with an amplitude of 0.011$^\circ$/day and a period of 70.8 years. Although this variation
is too small to account for the observed in the B-ring edge distortion, it does lead to
to an instantaneous mean motion at epoch (Day 2005-195) 
of 381.984$^\circ$/day, close to that derived from the JPL SPICE ephemeris. }

This trend in the pattern's orientation over the last few years 
can also be detected by comparing the VIMS occultation data
obtained in 2008 with various measurements of the B-ring
edge made by other instruments earlier in the Cassini Mission.
Data from 12 occultations of the radio signal from the
Cassini spacecraft (French et al. {\it in prep}) confirm that
in 2005 the $m=2$ radial excursions of the B-ring edge led Mimas
by $~70^\circ$ in phase (or $35^\circ$ in longitude). A similar result was
obtained by \citet{Spitale06}, based on Cassini imaging sequences in 2005.  
The pattern has therefore drifted backwards relative
to Mimas by $\sim55^\circ$ in longitude (or 110$^\circ$ in phase) between 2005 
and 2008, consistent with the drift rate obtained using the VIMS data alone.

A steady drift in the $m=2$ pattern relative to Mimas would be extremely
surprising, given that this pattern is supposed to be generated by 
gravitational perturbations from that 
moon. It seems more reasonable that the orientation of the $m=2$ pattern 
instead librates relative to the moon on a timescale that is
long compared to the Cassini mission to date. We therefore posit that the $m=2$ 
structure of the B ring edge can be described by the following equation:
\begin{equation}
r=r_o-A*\cos(2[\lambda-\lambda_{Mimas}-\phi_L*\sin(Lt-\theta_L)]),
\label{librateexp}
\end{equation}
where the last term in the cosine argument describes a slow libration of the longitude of
the minimum radius relative to Mimas with an amplitude of $\phi_L$ and 
a period of $2\pi/L$ (We shall see below that the amplitude $A$ also appears to 
be time-dependent). 

Estimating $\phi_L$ and $L$ from the Cassini VIMS data alone is difficult because
the libration period appears to be significantly longer than the observation arc. 
Earlier measurements of the B-ring edge orientation back to the Voyager missions
could be useful, but these data are rather sparse, so before we consider
using those data, it is useful to first place some constraints on the libration 
parameters using the Cassini VIMS data alone.

Since the pattern probably aligned with Mimas sometime during
2007, we can make the crude approximation that
during the entire Cassini mission to date $\sin(Lt-\theta_L) \simeq (Lt-\theta_L)$, in which case:
\begin{equation}
r\simeq r_o-A*\cos(2[\lambda-(n_{Mimas}+\phi_LL)t+\phi_L\theta_L-\lambda_{Mimas}(t=0)]),
\end{equation}
where $n_{Mimas}\simeq381.99^\circ/$day is the current mean motion of Mimas
.
In this case, the same basic procedure described above for 
finding the pattern speeds of the eccentric edges can be used to estimate 
the value of $n_{Mimas}+\phi_LL$ that best fits the data. Using all the 
VIMS occultation data the best fit pattern speed for the $m=2$ component
of the B-ring edge is found to be 381.945$^\circ$/day, or about 
0.045$^\circ$/day slower than Mimas' mean motion 
(see Figure~\ref{bedge2} bottom panel; using only the quality
data or including the radio science data does not change this result much). 
Thus the product $|\phi_LL|$ must be 0.045$^\circ$/day (where $\phi_L$ is 
measured in radians and $L$ is measured in degrees/day). 

To turn this estimate of $|\phi_LL|$ into a constraint on $L$, note that $\phi_L$
cannot have any value. First, $\phi_L$ cannot exceed
90$^\circ$ or else the pattern would be circulating instead of librating, which
seems unlikely. Also, $\phi_L$ is probably at least $35^\circ$ or 0.6 rad, 
since the pattern is offset by that much in 2005. These
considerations suggest that $L$ most likely lies somewhere
in the range between 0.03$^\circ$/day and 0.10$^\circ$/day, corresponding 
to a libration period of between 10 and 30 years.

Additional rough constraints on $\phi_L$ and $L$ can be obtained 
by recalling that analyses of the Voyager data showed that the $m=2$ pattern 
was aligned with Mimas in 1980-1981 \citep{Porco84}. This 
suggests that zero-crossings of the libration occurred in 
1980-1981 and 2007, which would imply
the libration period is a submultiple of 52 years, or
that $L$ is an integral multiple of $0.02^\circ$/day.  
The Cassini-based estimate of $|\phi_LL|$ and
the Voyager data can be satisfied if  $L \simeq 0.04^\circ$/day,  
$0.06^\circ$/day, $0.08^\circ$/day, etc. (with  $\phi_L
\simeq 60^\circ, 40^\circ$ and 30$^\circ$, etc. respectively).

Guided by these constraints, we performed a more comprehensive
investigation of the B ring edge using a combination of 
occultation data obtained over the last 30 years.
This combined data set includes: (1) the 48 VIMS occultation 
cuts listed in Table~\ref{obstab} (2) 12 occultation cuts from 2005 
obtained by the Cassini  radio-science experiment (RSS) reported 
in \citet{French10}  and kindly provided to us, 
(3) the Voyager  1 RSS occultation \citep{Tyler83},  (4) the
Voyager 2 PPS occultation \citep{Lane82, Esposito83, Esposito87}
(5) multiple data from the ground-based 28 Sagittarius occultations 
from July 1989 \citep{French93}, and (6)
occultations observed by the Hubble Space Telescope in 1991
\citep{Elliot93} and 1995 \citep{Bosh02}. All of the pre-Cassini data 
have been re-analyzed by  \citet{French10}. 

\begin{figure}
\centerline{\resizebox{5in}{!}{\includegraphics{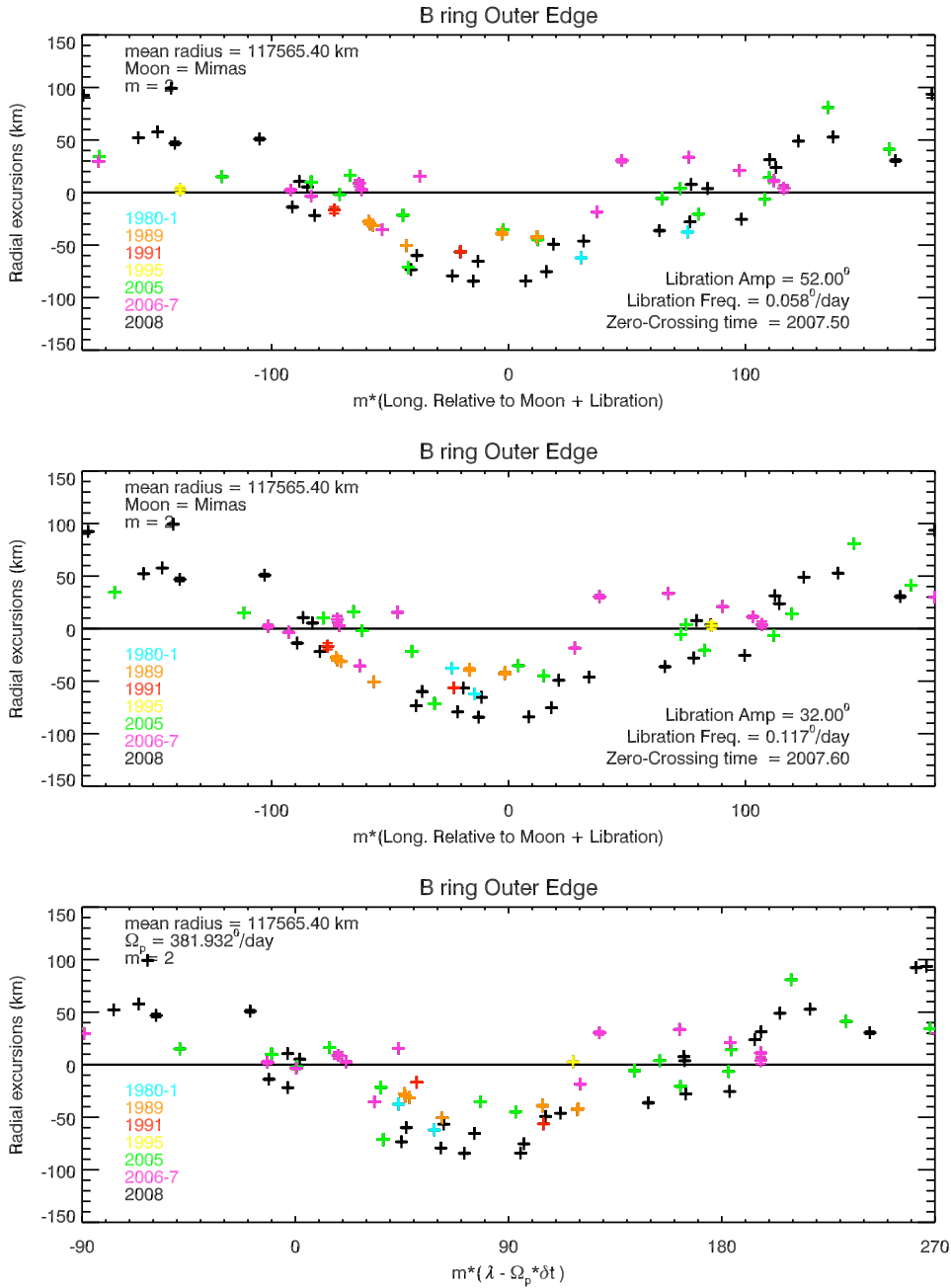}}}
\caption{Radial excursions of the B-ring edge, including data from
Cassini-VIMS and RSS experiments, Voyager observations
and various earth-based occultation observations. The top two panels
show the data as organized using a librating longitude system
with libration frequencies of 0.058$^\circ/$day and 0.117$^\circ/$day,
respectively. The bottom panel shows the data organized using a
steady drift rate of $\sim 0.05^\circ/$day slower than Mimas' mean motion
(as before, times are measured relative to an epoch of 2005-195T02:12:13.557 
or a Cassini spacecraft clock time of 150000000). The available data are clearly insufficient
to distinguish between these different models, given the very large residual scatter}
\label{bedge2o}
\end{figure}

To determine which combination of libration parameters could best
fit these data, we computed the Pearson's correlation coefficient
$\rho$ between the radial excursions and the parameter
$\cos(2[\lambda-\lambda_{Mimas}-\phi_L*\sin(Lt-\theta_L)])$
for different values of $\phi_L$, $L$ and $\theta_L$. This
correlation coefficient will be maximal with the model parameters that
best match the actual motion of the edge, so it provides
a convenient way to search the parameter space for likely
solutions. This method also has the advantage that it is
relatively insensitive to any time variability in the amplitude $A$
of the pattern (see below).

Based on this analysis, we found that solutions with 
$L<0.05^\circ$/day were strongly disfavored. This is primarily
because all of the pre-Cassini measurements show 
negative radial excursions, and there was no
way to align both the 28 Sgr and Voyager
measurements with the minima in the Cassini data
unless $L>0.05^\circ$/day. Furthermore, we found
that frequencies near 0.058$^\circ/$day and 0.117$^\circ/$day 
better organized the data than other frequencies. 
Note that both these frequencies are part of the
acceptable series of values that the above simplistic
analysis suggested would be compatible with both
the Voyager and Cassini data.

Unfortunately, the available data do not clearly
favor one of these two solutions for the B-ring edge over the
other. The top two panels in Figure~\ref{bedge2o} show 
the two best-fit solutions, with libration frequencies $0.058^\circ/$day and
0.117$^\circ/$day and libration amplitudes of $\phi_L=52^\circ$ and 
32$^\circ$ (in both cases the most recent
zero-crossing time was in mid-2007). The scatter of the data 
about the mean curves is not significantly different in these 
two cases, so both solutions are equally good in this respect. 
Worse, the earlier data are unable to rule out the possibility
that the $m=2$ pattern could be drifting backwards at a constant speed instead of librating.
The bottom panel of Figure~\ref{bedge2o} shows the data
plotted using the best-fit constant pattern speed, which turns out
to be 381.932$^\circ/$day, or about $0.05^\circ$/day slower
than Mimas' current mean motion. The scatter in the data for this solution 
is not much larger than it is for the librating solutions.

Future observations may eventually provide a way to
discriminate between these possible solutions. However, 
some of the difficulty in determining the correct model for
the motion of the $m=2$ pattern stems from the comparatively
large  scatter  in the measurements with respect to any of
the above solutions. In all cases, the average amplitude of the pattern is 
around 60 km, but  $rms$ residuals are roughly 20 km, which is 
much larger than the $\sim 1$ km  measurement errors in each of 
these data sets. This suggests that the B-ring edge is not just a 
fixed-amplitude $m=2$ pattern that librates relative to Mimas, 
but has a more complex shape, perhaps with additional 
perturbation modes.

\begin{figure}[tbp]
\centerline{\resizebox{5in}{!}{\includegraphics{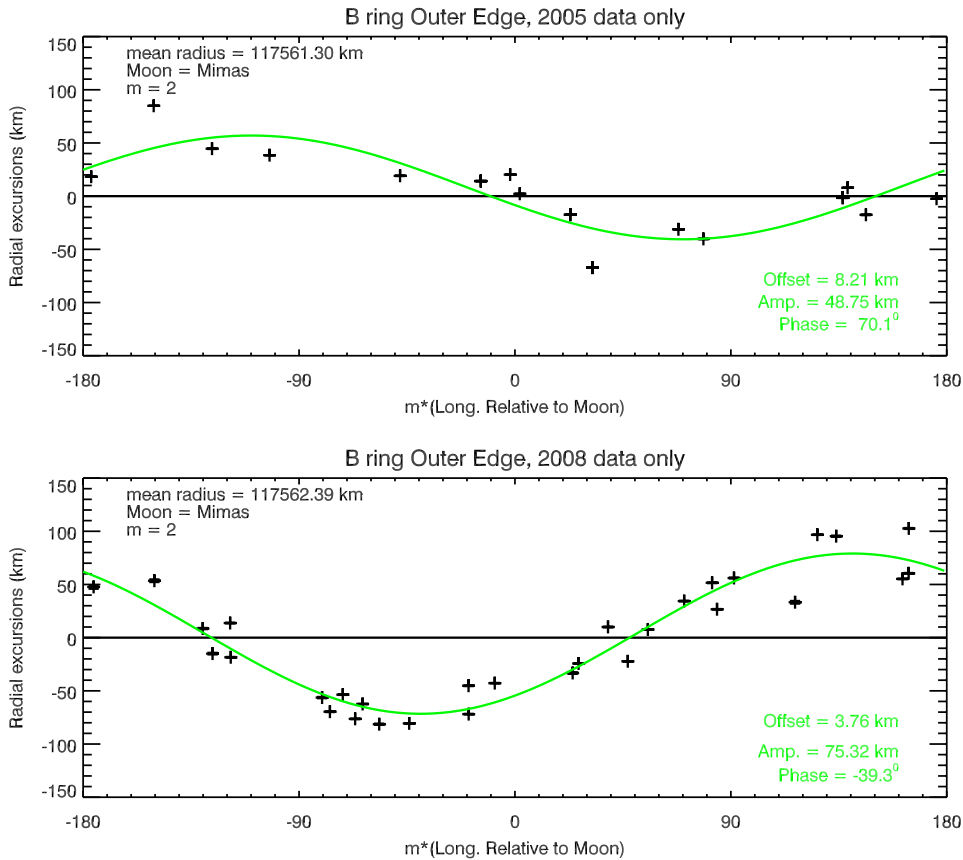}}}
\caption{Radial excursions of the B-ring edge measured by the VIMS and RSS 
experiments in 2005 (top) and by VIMS in 2008 (bottom). In both cases, the data
can be reasonably well fit by simple sine curves.}
\label{bedge2c}
\end{figure}

\begin{figure}[htbp]
\centerline{\resizebox{5in}{!}{\includegraphics{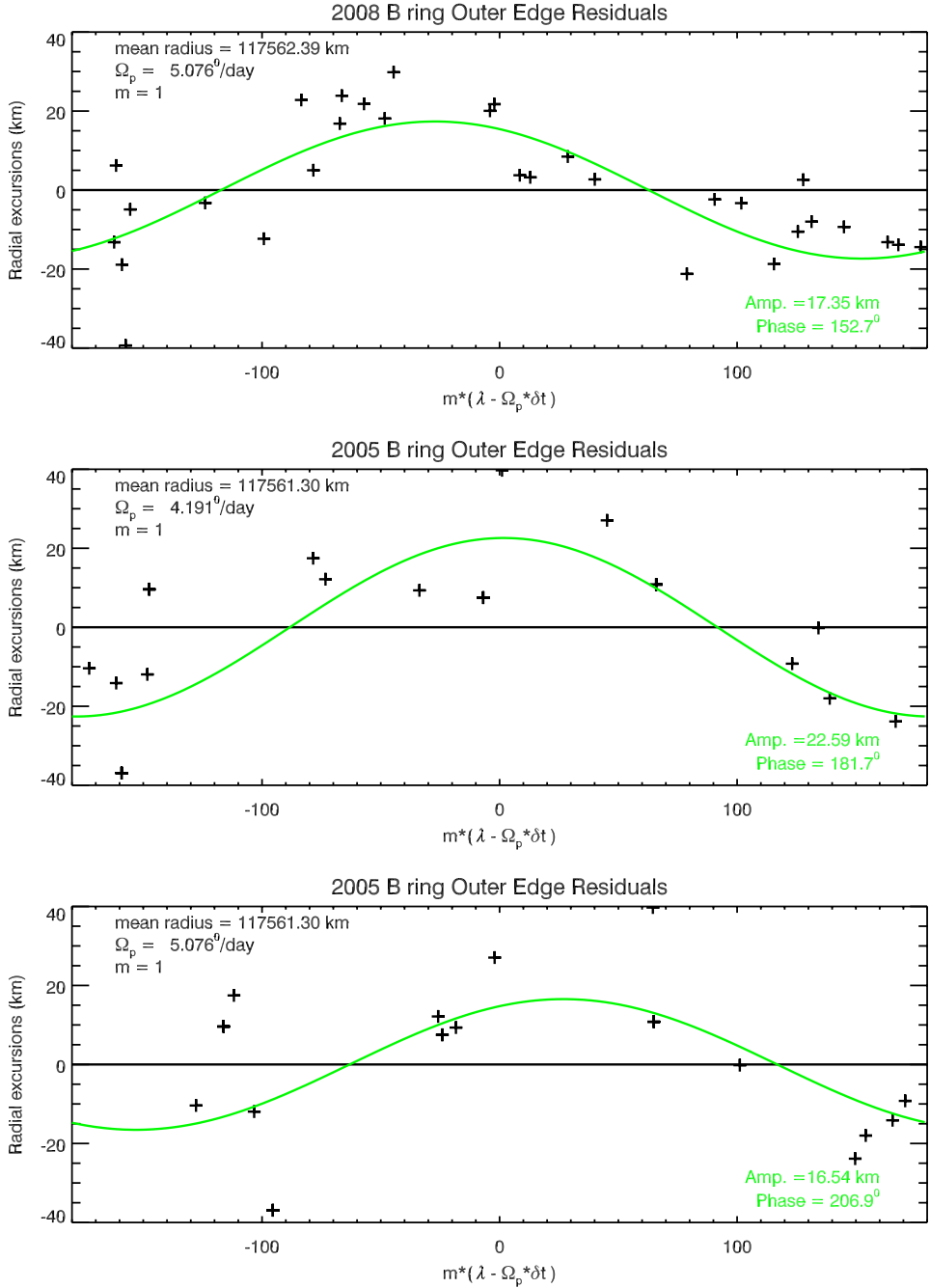}}}
\caption{Residual radial variations of the outer edge of the B ring, after removing
the $m=2$ mode pattern associated with the Mimas 2:1 resonance. The top and 
middle panels show the 2008 and 2005 residuals, respectively, each organized
using the best-fit pattern speed. Note the pattern speed for the 2008 data is
close to the apsidal precession rate at this location, while the 2005 data
is best fit by a 20\% slower pattern speed. The bottom panel shows
the 2005 data plotted using the same pattern speed as the 2008 data, to 
illustrate that the data are relatively well organized in this case as well. As before, all times are measured relative to an epoch of 2005-195T02:12:13.557 
or a Cassini spacecraft clock time of 150000000.}
\label{bedge1}
\end{figure}

To explore this possibility, let us take a closer look at two particular
subsets of the data: the VIMS data from 2008 and the combined VIMS and
RSS data from 2005. Each of these subsets consists of a 
reasonably large number of occultation cuts (31 and 16, respectively)
that were not only taken over a sufficiently broad range of longitudes 
relative to Mimas that we can estimate the shape of the $m=2$ pattern, but also
obtained in a  sufficiently short period of time that we can ignore
the libration of the pattern with respect to Mimas. 

The best-fit $m=2$ pattern for the 2008 data has an 
amplitude of  75.3 km and a phase of  -39.4$^\circ$ relative to Mimas
(i.e. the pattern lags Mimas by 19.7$^\circ$ in longitude), 
while the best-fit pattern for the 2005 data has an amplitude of 
48.7 km and a phase of +70.0$^\circ$. Thus the amplitude
of the $m=2$ pattern varies with time as well as the phase (see Figure~\ref{bedge2c}).
This is not entirely unexpected, given that a libration can
often be modeled as a combination of free and forced terms, 
which naturally leads to coupled variations in both the phase and the amplitude 
of the pattern (see Section 4 below). Indeed, some of the observed scatter in the data in 
Figure~\ref{bedge2o} is likely due to unmodeled changes in the 
amplitude of the $m=2$ pattern. 
Even so, the variations in the amplitude of the pattern cannot be the 
only source of scatter in these data, as each of the 2008 and 2005 
data sets alone show significant  scatter around the mean $m=2$ pattern, 
with individual residuals up to  $\sim$40 km (see Figure~\ref{bedge2c}).

A previous analysis of Cassini images by
\citet{Spitale06} suggested that additional perturbations, possibly $m=3$, might
influence the B-ring edge's shape. Motivated by this result, 
we took the 2008 and 2005 data sets, removed the best-fitting $m=2$ 
pattern from each of them, and fitted the residuals to 
$m=0$,1,3,4,5, and 6 patterns. For each non-zero value of $m$, we
examined a range of pattern speeds in the vicinity of the expected
speed of a ``normal mode" \citep{French91}  with this $m-$number:
\begin{equation}
\Omega_p=[(m-1)n+\dot{\varpi}]/m.
\end{equation}
For $m=0$, the pattern does not rotate, but the whole ring oscillates
radially at the epicyclic frequency $\kappa=n-\dot{\varpi}$, 
as observed for the Uranian $\gamma$ ring \citep{French86}.
As with eccentric features in the Cassini Division, we computed
the amplitudes:
\begin{equation}
\alpha_R=\frac{1}{n}\sum_{i=1}^n\delta r_i*\cos[m((1-\delta_{m0})\lambda_i-\Omega_p\delta t_i)]
\end{equation}
\begin{equation}
\alpha_I=\frac{1}{n}\sum_{i=1}^n\delta r_i*\sin[m((1-\delta_{m0})\lambda_i-\Omega_p\delta t_i)]
\end{equation}
where $\delta r_i$ are the residual radial variations in the edge positions
after removing the $m=2$ pattern. As before, if there is a pattern in these
radius measurements with a given $m$ and $\Omega_p$, then
$\sqrt{\alpha_R^2+\alpha_I^2}$ should have a maximum at the appropriate
values for those parameters.

For both the 2005 and 2008 data sets, the clearest maxima were obtained  with  
$m=1$. Indeed, as shown in Figure~\ref{bedge1}, 
the residuals in the 2008 data are reasonably well fit by a precessing 
Keplerian ellipse with an amplitude of about 18 km and a pattern speed 
of 5.08$^\circ$/day (with an uncertainty of $\sim0.05^\circ$/day determined
by the $\sim$6-month observing arc). 
This pattern speed  is very similar to the expected apsidal precession 
rate of 5.061$^\circ$/day for a Keplerian orbit at this radius.
For the 2005 data, the best-fit solution has a slightly
larger amplitude of about 23 km and the best-fit pattern speed of 4.19$^\circ$/day, 
17\% slower than the predicted value of $\dot{\varpi}$. However, given the
short time-baseline covered by the 2005 data and the limited number of
data points available, a pattern speed of $5.08^\circ$/day
(consistent with the 2008 observations) can also organize the 2005 data 
reasonably well (see bottom panel of Figure~\ref{bedge1}). 
An attempt to fit a single, coherent $m=1$ perturbation to both data sets
yields multiple equal-quality fits at an array of pattern speeds separated
by $360^\circ/$3 years $\simeq 0.3^\circ$/day. The available data are therefore
too sparse to permit us to derive a more precise model of this perturbation
or an accurate measurement of the pattern speed. However, the existence
of similar $m=1$ patterns in both the 2005 and 2008 data sets strongly suggests 
that this is a ``permanent'' feature of the B-ring edge.

The radial variations in the location of the B-ring edge therefore have at least 
two components: an $m=2$ perturbation 
forced by the strong Mimas 2:1 Inner Lindblad Resonance, with an average radial
amplitude of $\sim$60 km and an orientation which librates (or circulates) with respect to the
direction to Mimas, and an $m=1$ Keplerian ellipse with an amplitude of $\sim 20$ km which freely precesses under the influence of Saturn's oblate figure at  roughly $5^\circ$/day. Even this
rather complicated model provides a relatively poor fit, in comparison with those
for the eccentric edges in the Cassini Division, with $rms$ residuals of 10-20 km. We suspect
that this is due to deficiencies in our model of the $m=2$ perturbation, whose libration 
is still not accurately modeled (e.g. we have not yet implemented a model for the amplitude variations), 
but it is also possible that additional  perturbations are present. 
Efforts to derive more detailed and accurate
models of the B-ring edge are underway and will be presented in future work, 
but already the preliminary results presented here hint at a close connection
between the B-ring edge and the Cassini-Division gaps.

\section{Theoretical expectations for the B-ring outer edge}

Before delving into the possible connections between the
B-ring edge and the Cassini-Division gaps, it is useful to first  
take a closer look at the complex behavior of the
B-ring's outer edge. In particular, the above data
allow several different solutions for the motion of the $m=2$
component on this edge, all of which have the $m=2$ pattern
drift or move relative to Mimas. We would like to
establish whether any of these solutions
are plausible in terms of the local dynamical environment.
Since a detailed model along the lines of 
those developed in ~\citet{BGT82} and \citet{Hahn09}
is beyond the scope of this paper, we will instead 
examine the behavior of isolated ring particle orbits 
in the vicinity of the Mimas 2:1 inner Lindblad resonance (ILR). 
While not conclusive, these simpler calculations
do at least demonstrate that the observed
variations in the amplitude and orientation of the $m=2$
pattern are not wildly inconsistent with  theoretical expectations.

The following discussion is based on that in Chapter 8
of Murray and Dermott (1999), but couched in terms of the physical
coordinates, $r$ and $\lambda$ and the standard orbital elements, $a,
e$ and $\varpi$, rather than a Hamiltonian formalism.

At a first-order ILR, the resonant argument is given by 
\begin{equation}
\varphi_{\rm ILR} = (m-1)\lambda + \varpi - m\lambda_s
\label{resarg}
\end{equation}
\noindent
where $m$ is a positive integer, $\lambda$ and $\lambda_s$ refer to
the mean longitudes of the test particle and satellite, respectively,
and $\varpi$ is the longitude of pericenter of the test particle's
orbit. At the so-called exact resonance,
\begin{equation}
\frac{d\varphi_{\rm ILR}}{dt} = (m-1)n + \dpi - m n_s = 0
\label{exact}
\end{equation}
\noindent
where $n$ is the orbital mean motion and $\dpi$ is the apsidal
precession rate of the test particle due to non-resonant perturbations
(chiefly the planet's oblateness). For a 2:1 ILR, $m=2$ and $\nres =
2n_s - \dpi.$

An ensemble of test particles which share common values of semimajor
axis $a$, eccentricity $e$ and $\varphi_{\rm ILR}$, but different instantaneous
values of $\lambda$ and $\varpi$, will define a streamline given by
\begin{equation}
r(\lambda,t) \approx a\left[1 - e \cos(\lambda - \varpi)\right],
\label{ellipse}
\end{equation}
\noindent or in terms of $\varphi_{\rm ILR}$:
\begin{equation}
r(\lambda,t) \approx a\left[1 - e \cos\left(m(\lambda-\lambda_s) - 
\varphi_{\rm ILR}\right)\right].
\label{streaml}
\end{equation}
\noindent 
This is the same as Eq.(\ref{librateexp}) above, with $A = ae$, $r_0' = a$
and
\begin{equation}
\varphi_{\rm ILR} = 2\phi_L\sin(Lt-\theta_L).
\label{philib}
\end{equation}

Following Murray and Dermott (1999, see their Eq. (8.26) and Table 8.5), the
corresponding time-averaged disturbing function of the
satellite (to lowest order in eccentricity and for zero
inclination) is given by
\begin{equation}
\mathcal{R}_{\rm ILR} = \frac{Gm_s}{a_s}f(\alpha)e\cos\varphi_{\rm ILR}
\label{distfn}
\end{equation}
\noindent 
where $m_s$ is the mass of the satellite, $\alpha = a/a_s$ and for
$m=2$ the function $f(\alpha) = -0.75/\alpha$ comes from evaluating the disturbing function. 
The Lagrange equations
which describe the resulting perturbations in the test particle's
orbital elements are:
\begin{equation}
\frac{dn}{dt} = -3(m-1)\beta n^2 e\sin\varphi_{\rm ILR},
\label{dndt}
\end{equation}
\begin{equation}
\frac{de}{dt} = -\beta n \sin\varphi_{\rm ILR},
\label{dedt}
\end{equation}
\begin{equation}
\frac{d\varpi}{dt} = -\beta n e^{-1}\cos\varphi_{\rm ILR} + \dpi,
\label{dpidt}
\end{equation}
\begin{equation}
\frac{d\epsilon}{dt} = -\frac{1}{2}\beta n e\cos\varphi_{\rm ILR},
\label{depsdt}
\end{equation}
\noindent
where we have introduced the dimensionless resonance strength
$\beta = -(m_s/M_S)\alpha f(\alpha)$, with $M_S$ the mass of the
planet. The quantity $\epsilon$ is the longitude at epoch, defined
by the usual expression $\lambda = \epsilon + nt$.  These are
equivalent to Eqns. (8.28--8.31) of Murray and Dermott (1999), except that our
definition of the resonant argument in Eq.~(\ref{resarg}) is opposite
in sign to theirs. Combining Eqns.~(\ref{dndt}, \ref{dpidt} and
\ref{depsdt}), we can derive an expression for the rate of change of
$\varphi_{\rm ILR}$
\begin{equation}
\frac{d\varphi_{\rm ILR}}{dt} = (m-1)n - mn_s + \dpi - \left[\frac{\beta n}{e}
+ \frac{(m-1)\beta n e}{2}\right]\cos\varphi_{\rm ILR}.
\label{dphidta}
\end{equation}
\noindent
For small values of $e$, we can safely neglect the last term
in brackets (which arises from Eq.~(\ref{depsdt})).

From Eqns.~(\ref{dndt} and \ref{dedt}), we see that $n$ and $e$ must vary
in phase, with
\begin{equation}
\frac{dn}{de} = 3(m-1)ne,
\label{dnde}
\end{equation}
\noindent
so that we may write, to the lowest order in $e$
\begin{equation}
n \approx n_0 + \frac{3}{2}(m-1)n_0 e^2,
\label{ne}
\end{equation}
\noindent
where $n_0$ is a constant. Substituting this expression for $n$ into
Eq.~(\ref{dphidta}), and introducing the constant parameter $\nu =
(m-1)n_0 - mn_s + \dpi$, we have our final equation for the resonant
variable, correct to the lowest order in $e$:
\begin{equation}
\frac{d\varphi_{\rm ILR}}{dt} = \nu + \frac{3}{2}(m-1)^2n_0 e^2 - 
\frac{\beta n_0}{e}\cos\varphi_{\rm ILR}.
\label{dphidtb}
\end{equation}
\noindent
The frequency parameter $\nu$ is best thought of as a measure of the 
distance from exact resonance,
\begin{equation}
\nu \approx -\frac{3}{2}(m-1)\nres\left(\frac{a-\ares}{\ares}\right)
\label{nudef}
\end{equation}
\noindent
where $\ares$ and $\nres$ are the semimajor axis and unperturbed mean
motion at exact resonance, defined by Eq.~(\ref{exact}).  Note that
$\nu > 0$ for orbits {\it interior} to $\ares$.


Equilibrium solutions   to Eqns.~(\ref{dedt} and \ref{dphidtb}) occur where
$e=e_0$ and $ \varphi_{\rm ILR}=\varphi_0$, where $e_0$ and $\varphi_0$ are
constants in time. Such solutions only exist
for $\varphi_0=0$ or $\varphi_0=\pi$, and where the eccentricity
satisfies the cubic equation:
\begin{equation}
\nu = \frac{\beta n_0}{e_0}\cos\varphi_0 - \frac{3}{2}(m-1)^2n_0 e_0^2. 
\label{eforc}
\end{equation}
\noindent
For positive values of $\nu$ (i.e. interior to $a_{\rm res}$) only a single
solution exists, with $\varphi_0=0$. But for negative values of $\nu$
(i.e. exterior to $a_{\rm res}$) up to three branches of solutions exist,
labelled A, B and C by Murray and Dermott (1999, see their
Fig.~8.9). Solution A, with $\varphi_0=0$, is an extension of the
equilibrium solution for $\nu>0$.  Solutions B and C have
$\varphi_0=\pi$, with $e_B\leq e_C\leq e_A$. Solution C is an unstable
equilibrium, and thus of little physical interest, but solutions A and
B are both stable. Branches B and C merge at a bifurcation point,
where $d\nu/de_0=0$,
\begin{equation}
\nu_c = -\frac{3}{2}n_0 \left[3(m-1)^2\beta^2\right]^{1/3},
\label{nucrit}
\end{equation}
\noindent and
\begin{equation}
e_c = \left[\frac{\beta}{3(m-1)^2}\right]^{1/3}, 
\label{ecrit}
\end{equation}
\noindent
and cease to exist for larger values of $\nu$.

For $e<<e_c$, the equilibrium, or {\it forced} eccentricity is given
approximately by
\begin{equation}
e_0 \approx \beta\frac{n_0}{\nu} \cos\varphi_0,
\label{esmall}
\end{equation}
\noindent 
with $\varphi_0=0$ for $\nu>0$ (solution A) and $\varphi_0=\pi$ for
$\nu<0$ (solution B).  For $e>>e_c$, $\varphi_0=0$ (solution A) or
$\pi$ (solution C) and
\begin{equation}
e_0 \approx \left[\frac{2|\nu|}{3n_0(m-1)^2}\right]^{1/2}.
\label{ebiga}
\end{equation}
\noindent
At $\nu = 0$, $e_0 = 2^{1/3}e_c$.

In the context of planetary rings, it is usually assumed that we are
in the `small-$e$' regime, with $e<<e_c$ and a resonantly-forced
eccentricity and phase given by Eq.~(\ref{esmall}) (cf. Murray and
Dermott 1999, Eq.~(10.21) and Fig.~10.9).

If we now substitute suitable numerical values for the Mimas 2:1 ILR,
we find that $a_{\rm res} = 117,553.65$~km, $\dpi =
5.0613^\circ$/day, $\nres = 2n_s - \dpi =
758.9277^\circ$/day, $\alpha \approx 0.6336$, $m_s/M$ = $6.60*10^{-8}$
\citep{Jacobson06}, and $\beta = 4.95*10^{-8}$.  The corresponding
eccentricity and frequency at the bifurcation point are $e_c =
0.00255$ and $\nu_c = -0.0221^\circ$/day. From Eq.~(\ref{nudef})
we have that
\begin{equation}
\nu \approx -0.0097^\circ/{\rm day}\left(\frac{a-\ares}{1~{\rm km}}\right),
\label{nudeg}
\end{equation}
\noindent
so that the bifurcation point occurs at $a-a_{\rm res} = +2.3$~km with a
forced amplitude, $ae_c = 299$~km.

Since the maximum radial amplitude observed at the edge of the B ring,
$A \simeq 75~$km, or one-quarter of the critical value, we may
conclude that the streamlines are indeed in the `small-$e$' regime.
We can then use Eq.~(\ref{esmall}) to estimate the effective value of
$\nu$ for streamlines at the edge of the B ring. If we adopt an
average forced eccentricity of ($0.50\pm0.12)*10^{-3}$ (i.e. $A =
60\pm15$~km), and note that $\varphi$ appears to librate around $0$
rather than $\pi$, then we find
\begin{equation}
\nu_{\rm eff} \approx \frac{\beta n_0}{e_0}\cos\varphi_0 = 
+0.075\pm0.019^\circ/{\rm day}
\label{nueff}
\end{equation}
\noindent 
with an effective distance from exact resonance of $a-\ares =
-8\pm2$~km. This may be compared with the observed mean location of the
edge of 117,565.4~km, or $a-a_{res} = +12$~km.  The B ring edge thus
{\it behaves} as if it were located $5-10$ km interior to the 2:1
resonance, whereas it is {\it actually} located $\sim12$~km exterior
to the resonance.  A similar conclusion was reached by \citet{Porco84}
in their study of this edge using Voyager observations, though
their value of $a-\ares$ was less accurate due to uncertainties at
that time in the absolute radius scale of the rings.

However, the classical `resonance width'  $w_{res}$ --- the range in semimajor axis
over which test particle orbits on opposite sides of exact resonance
will overlap due to their $180^\circ$ difference in $\varphi_0$ --- is
given by substituting Equation~\ref{nudef} into Equation~\ref{esmall} 
and solving for $ae_o=a-a_{res}=w_{res}/2$
(cf. Murray and Dermott 1999, Eq.~(10.23))
\begin{equation}
w_{\rm res} \approx 2\ares\left[\frac{2\beta}{3(m-1)}\right]^{1/2} = 43~\rm{km}.
\label{width}
\end{equation}
\noindent
(Note that this is actually $1/2$ of Murray and Dermott's expression, which includes the
radial excursions due to the forced eccentricities.) Within $\sim20$~km of exact resonance,
therefore, we cannot expect the above test particle model to give
realistic estimates of $e_0$ as a function of $a$.  In an actual
planetary ring, the situation will be further complicated by
gravitational and collisional interactions between the ring particles,
which will act to prevent the streamline crossing predicted by our
simple test-particle model within $\pm w_{\rm res}/2$ of $a_{\rm res}$. Such a model
has recently been developed for application to the A and B ring edges
by \citet{Hahn09}, based on earlier work by \citet{BGT82}.


Although dissipative collisions within a real ring might be expected
to rapidly damp any motion relative to the equilibrium solutions
discussed above, our observations of the B ring edge strongly suggest
either that the resonance angle $\varphi_{\rm ILR}$ is librating about an
equilibrium value close to $0$, or that it is circulating in a
retrograde direction (i.e. $\langle d\varphi_{\rm ILR}/dt \rangle < 0$). The
equations of motion, (\ref{dndt} -- \ref{depsdt}) admit of both
finite-amplitude oscillations in $\varphi_{\rm ILR}$ and of circulating
solutions. For a series of phase portraits of such librations the
interested reader is referred to Fig.~8.10 in Murray and
Dermott (1999). (Their dimensionless parameter $\delta$ is equal to
$3\nu/\nu_c$, while their amplitude parameter $\Phi =
\frac{1}{2}(e/e_c)^2$.)


In the small-$e$ limit, (i.e. $e <<
e_c$) Eq.~(\ref{dphidtb}) reduces to
\begin{equation}
\frac{d\varphi_{\rm ILR}}{dt} \approx \nu - \frac{\beta n_0}{e}\cos\varphi_{\rm ILR},
\label{dphidtc}
\end{equation}
\noindent
while the variation in $e$ is given by Eq.~(\ref{dedt}), where we may
set $n \approx n_0$. In this limit, the variations in $\varphi_{\rm ILR}$ are
dominated by the resonant effects on $\varpi$.  These coupled
equations are most readily solved by introducing the conjugate
variables
\begin{equation}
h = e \cos\varphi_{\rm ILR}\mbox{, }  k = e \sin \varphi_{\rm ILR},
\label{hkdef}
\end{equation}
\noindent
in terms of which Eqns.~(\ref{dedt}) and (\ref{dphidtc}) become
\begin{equation}
\frac{dh}{dt} = -\nu k,  
\end{equation}
\begin{equation}
\frac{dk}{dt} = \nu h - \beta n_0.
\label{dhkdt}
\end{equation}
\noindent
The solution follows trivially:
\begin{equation}
h(t) = \frac{\beta n_0}{\nu} + e_f \cos(\nu t - \theta),
\end{equation}
\begin{equation}
k(t) = e_f \sin(\nu t - \theta).
\label{hksol}
\end{equation}
\noindent
Recall that $\beta n/|\nu|$ is the forced eccentricity, $e_0$ from
Eq.~(\ref{esmall}) above. The {\it free eccentricity}, $e_f$ and phase
angle $\theta$ are arbitrary constants of the motion, set by the
initial conditions.  In the $h,k$ plane, the motion is in a circle of
radius $e_f$ about the fixed point, ($\pm e_0$, 0). Interior to the
resonance ($\nu > 0$, equilibrium branch A) the motion is
counterclockwise about ($+e_0$, 0), while exterior to the resonance
($\nu < 0$, branch B) the motion is clockwise about ($-e_0$, 0). In
both cases, the angular frequency is simply equal to $\nu$, given by
Eq.~(\ref{nudef}) above.

If $e_f < e_0$, the solution describes a librational motion of
$\varphi$ about either 0 (for $\nu > 0$) or $\pi$ (for $\nu < 0$).
The maximum libration amplitude is given by $\varphi_{\rm max} =
\sin^{-1}(e_f/e_0)$, while the instantaneous eccentricity is equal to
\begin{equation}
e(t) = \sqrt{h^2+k^2} = \left[e_0^2 + e_f^2 + 2e_o e_f\cos(\nu t -
\theta)\right]^{1/2}.
\label{esol}
\end{equation}
\noindent
For $e_f > e_0$, the angle $\varphi_{\rm ILR}$ circulates continuously through
$2\pi$ radians while $e$ oscillates between $e_f-e_0$ and $e_f+e_0$.
Example trajectories for both cases are shown in Fig.~8.11 of Murray
and Dermott (1999).

Although the form of the small-$e$ librations motivated our choice
of the model used to fit the B ring edge above, we note that even this
simple test-particle model implies (i) that the amplitude of the edge,
$A = ae$ should oscillate as the edge librates or circulates, and (ii)
that the variation in the resonant angle, $\varphi_{\rm ILR}$ will not be
sinusoidal or linear, unless $e_f << e_0$ or $e_f >>e_o$. The substantial variations
observed in $A$ and $\varphi_{\rm ILR}$ over the course of the Cassini mission
to date suggest that the latter is unlikely to be true. A more
sophisticated model might therefore  be based on the small-$e$ solution above.

Despite these limitations, we note that the best-fitting libration
frequencies of $0.058^\circ$ and $0.117^\circ$/day
found above are generally consistent with our theoretical estimate of
$\nu = 0.075\pm0.019^\circ$/day in Eq.~(\ref{nueff}), based on
equating the observed amplitude to the forced eccentricity. 

If, on the other hand, the B ring edge is circulating, then the
best-fitting drift rate of $\phi_L L = -0.05^\circ/$day
(cf. Fig.~8) implies that $d\varphi_{\rm ILR}/dt = 2\phi_L L \approx
-0.10^\circ$/day, from Eq.~(\ref{philib}), or $a-\ares =
+10$~km, which agrees fairly well with the observed mean radius of the edge.  Thus
both librating and circulating small-$e$ models seem to be
quantitatively consistent with our observations of the B ring edge,
despite their obvious limitations noted above.

For completeness, we did examine the 
large-$e$ case, but these calculations indicate that a large-$e$ libration 
is less compatible with the observed motion of the B ring edge. Not only 
does the predicted forced amplitude, $A = ae_0$ exceed 750~km 
only 5~km outside the resonant radius --- ten times the maximum 
observed value ---, but the libration frequency is quite slow, reaching only 
$0.024^\circ$/day at $a-\ares = 5$~km and $0.033^\circ$/day at 20~km. Such low
libration frequencies are strongly disfavored by the available data.

\section{A possible explanation for the location of the Cassini Division gaps}

The above analysis indicates that the structure of the Cassini Division may be
more regular than it appears at first glance. Certainly, the Barnard Gap is no
longer a mystery, as it seems to be held open by the 5:4 mean motion resonance
with Prometheus. Furthermore, the remaining gaps seem to form a pattern. 
The Herschel, Russell, Jeffreys, Kuiper, Laplace and Bessel gaps all seem
to have an eccentric inner edge and most of these also have a circular outer edge. 
(The Huygens, Herschel and Laplace gaps do not have perfectly circular outer edges, but these
are also the gaps which contain dense ringlets, which may complicate the situation.)  
Furthermore, the 6 eccentric gap edges seem to  define a series of evenly spaced pattern 
speeds with a characteristic spacing of about  $0.06^\circ$/day. Even the Huygens  ringlet 
and the $m=1$ component of the B ring edge seem to fall in this sequence. Finally, there is the suggestive coincidence that the $m=2$ pattern on the B ring edge could be librating with an angular frequency
not too dissimilar from 0.06$^\circ/$day (although we must caution that other 
solutions are possible).

Based on these findings, we have developed a novel explanation
for the location of the gaps in the Cassini Division. Just as the inner
edges of the Huygens and Barnard Gaps are established by resonances
with saturnian satellites (Mimas 2:1 and Prometheus 5:4, respectively), 
the inner edges of the other gaps are maintained by resonances which 
involve the edge of the B ring. In the following sections, we demonstrate that
the gravitational perturbations from the radial excursions of the B-ring edge,
together with perturbations from Mimas, can give rise to terms in the equations
of motion that would support  the formation of eccentric edges at their observed
locations in the Cassini Division.   

Note that for the purposes of this theoretical study, we will assume 
that the orientation of the $m=2$ pattern on the B-ring edge librates 
with a frequency of $\sim0.06^\circ$/day, which matches the typical spacing between
the pattern speeds of eccentric features in the Cassini Division.
This particular solution is clearly the one where
the B-ring-edge patterns would be most likely to generate a series of
structures like the Cassini Division gaps. Thus an analysis of 
this case will establish whether such a model has any hope 
of working. Of course, our current understanding of the data
admits the possibility of different libration frequencies or even
circulation of the $m=2$ pattern. If one of these alternate solutions
turns out to be correct, then the connection between the B-ring edge
and the Cassini Division may still exist but be more complicated, 
and the simpler analysis presented here could still help clarify this
relationship.

\subsection{Qualitative Frequency Studies}

To determine if any interesting resonances
in the Cassini Division could be generated by the B-ring 
edge, one must write down the gravitational potential associated
with the observed structure of the B-ring edge, compute
the appropriate terms in the disturbing function, and determine
if they could drive structures
like those seen in the various gap edges. 
In this case, however, the data described above already provide
hints of  what terms in the disturbing function could be
involved in generating the observed edges. Therefore, prior to 
exploring the dynamical environment of the 
Cassini Division in detail, we will first take a more qualitative 
look at the situation in order
to clarify which terms in the gravitational potentials
are likely to be relevant.  

The $m=1$ patterns on the relevant Cassini Division edges
move at speeds close to the expected local apsidal precession
rates given Saturn's oblateness (See Table~\ref{ecctab}). Thus
each edge can be thought of as a collection of particles 
on freely-precessing eccentric orbits whose pericenters are all aligned
to produce a coherent structure. 
Assuming the orientation of the $m=2$ pattern in the 
B-ring edge librates at $\sim0.06^\circ/$day, this alignment
in pericenters would appear to occur at places where the
apsidal precession rate $\dot{\varpi}$ has the  following values:
\begin{equation}
\dot{\varpi}=\dot{\varpi}_B-jL
\label{cond1}
\end{equation}
where $\dot{\varpi}_B$ is the apsidal precession rate of the B-ring edge
(also the pattern speed of the $m=1$ component on that edge), 
$j$ is an integer and $L$ is the libration frequency of the $m=2$
pattern in the B-ring edge. Since $\dot{\varpi}$ as a function of $a$
is determined primarily by the higher-order components in Saturn's
gravitational field, Equation~\ref{cond1} implicitly specifies the locations of a
series of regularly-spaced resonances. 
At these locations, the following
quantity is approximately constant:
\begin{equation}
\varphi=\varpi-\varpi_B+jLt
\end{equation}
We propose that there is a term in the equations of
motion that tries to maintain $\varphi$ near
some value $\varphi_o$. Such a term will act
to align orbital pericenters and produce 
a coherent structure on each edge that is 
stable against small perturbations. In order for this to work,
the equation of motion of this resonant argument needs
to have a term of the form:
\begin{equation}
\frac{d^2\varphi}{dt^2}\simeq -f^2_o\sin(\varphi-\varphi_o).
\end{equation}
In this particular situation, the above expression can be
re-written as (assuming $\dot{\varpi}_B$ and $L$ are constant):
\begin{equation}
\frac{d^2\varpi}{dt^2}\simeq-f^2_o*\sin(\varpi-\varpi_B+jLt-\varphi_o)
\label{eqho}
\end{equation}
Provided such a term exists in the equations of motion, then the pericenters
of individual particle orbits could become aligned, forming
a coherent $m=1$ pattern that moves around the planet at
the local precession rate established by Saturn's
oblateness, consistent with the observations. 

Lagrange's Planetary Equations relate time derivatives
in orbital elements like $\varpi$ to derivatives of
the disturbing function (e.g. $d\varpi/dt\propto d\mathcal{R}/de$). 
Thus, in order to determine
if $\varphi=\varpi-\varpi_B+jLt$ is a proper resonant
argument that can produce a coherent structure, we need to find a term or
a combination of terms in the disturbing
function that are proportional to 
$\sin(\varpi-\varpi_B+jLt)$.

The question now is whether such terms are likely to
appear in the disturbing function generated by the B-ring 
edge. We can model the B ring edge as a collection of 
mass ribbons, each of which has the characteristic shape
\begin{equation}
r=r_o-d_1\cos(\phi-\varpi_B)-\tilde{d}_2\cos(2[\phi-\lambda_{Mimas}-\tilde{\phi}_L])
\label{bedgeq}
\end{equation}
where $\phi$ is the azimuthal angle. $\tilde{\phi}_L=\phi_L*\sin(Lt-\theta_L)$
is the time-variable azimuthal offset due to the edge's libration (cf. Equation~\ref{librateexp} above). Note the term $\tilde{d}_2$ 
may also have a time-variable component
($\tilde{d}_2=d_2[1\pm\delta^d_2\cos(Lt-\theta_L)]$) because the amplitude
of the $m=2$ pattern is apt to change over the course of the libration cycle.
In fact, as discussed in Section 4 above, the librations in the amplitude and the phase are likely to be coupled
in such a way that neither oscillation is purely sinusoidal, but for the purposes
of this analysis we will ignore such complications.

Such mass ribbons will produce a relatively complicated potential, 
but in general we expect terms in the disturbing function to contain
second-order terms of the following form
\begin{equation}
\mathcal{R}_{kl}\propto 
d_1^{|k|}\tilde{d}_2^{|l|}\cos(k(\lambda-\varpi_B)+l[2\lambda-2\lambda_{Mimas}-2\tilde{\phi}_L]),
\end{equation}
where $k$ and $l$ are various integers and $\lambda$ is the perturbed particle's
mean longitude (assuming the particle's orbit is nearly circular). We can quickly see that
any term of this form that contains $\tilde{\phi}_L$ will also 
contain $\lambda_{Mimas}$, so no term
of this form will have the form required to produce something like
Equation~\ref{eqho}. This suggests that the B ring edge alone 
will be unable to produce the resonances required to
explain the eccentric edges in the Cassini Division.

All is not lost, however, because there is a possibility that
the perturbations from the B ring edge, acting together 
with perturbations from the nearby and very powerful
Mimas 2:1 Inner Lindblad Resonance, can generate the desired
terms in the disturbing function through a sort of three-body resonance.
In such a situation, the time-derivatives of orbital elements 
can be proportional to the products of disturbing functions, which
means we will get products of the $\mathcal{R}_{kl}$ above
with the term responsible for the Mimas 2:1 resonance:
\begin{equation}
\mathcal{R}_{Mimas}\propto m_M\cos(\lambda+\varpi-2\lambda_{Mimas}),
\end{equation}
where $m_M$ is the mass of Mimas. This yields terms of the following form:
\begin{equation}
\mathcal{R}'_{k,l}\propto m_Md_1^{|k|}\tilde{d}_2^{|l|}\sin((\lambda+\varpi-2\lambda_{Mimas})
+k(\lambda-\varpi_B)+l[2\lambda-2\lambda_{Mimas}-2\tilde{\phi}_L])
\end{equation}

\begin{figure}[htbp]
\centerline{\resizebox{5in}{!}{\includegraphics{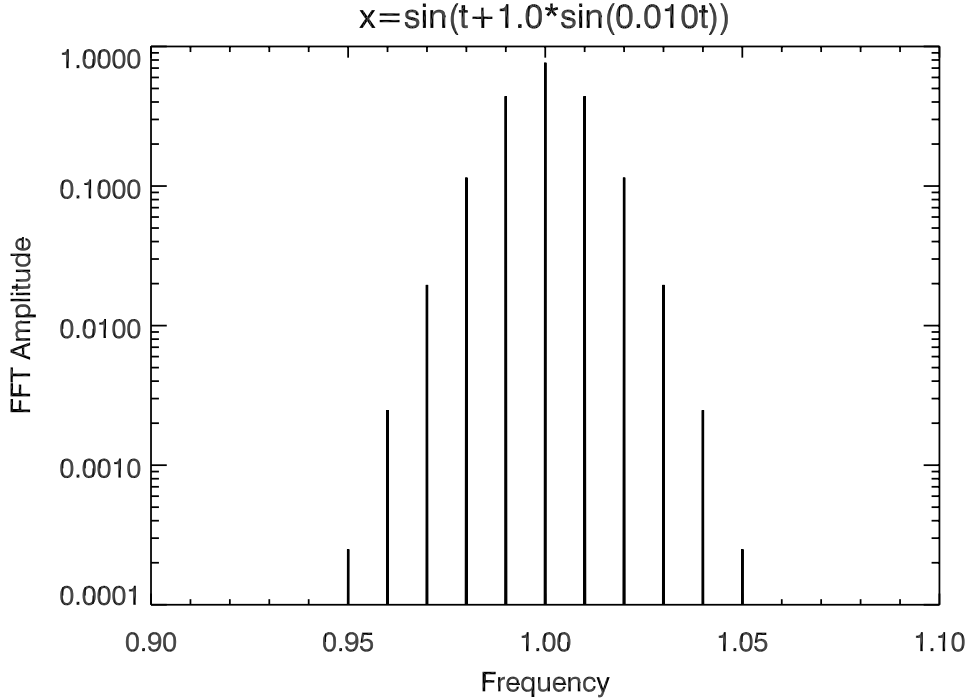}}}
\caption{Fourier transform of the function $x=\sin(t+\sin(.01t))$, showing the
multiple frequency components with comparable amplitudes.}
\label{fftlib}
\end{figure}

When $k=1$ and $l=-1$ the term has the promising form:
\begin{equation}
\mathcal{R}'_{1,-1} \propto m_Md_1\tilde{d}_2\sin(\varpi-\varpi_B+2\tilde{\phi}_L),
\label{term1}
\end{equation}
or, equivalently
\begin{equation}
\mathcal{R}'_{1,-1} \propto m_Md_1\tilde{d}_2\sin(\varpi-\varpi_B+2{\phi}_L\sin(Lt-\theta_L)).
\end{equation}
This term is  proportional to something like 
$\sin(A*t+B*\sin(C*t))$. Note in particular that  in this
case $\phi_L \simeq 50^\circ$ or nearly 1 radian (cf Fig.~\ref{bedge2o}), so the coefficient
$B$ is actually quite large.  Figure~\ref{fftlib}
shows a fourier transform of a function of the form 
$\sin(t+\sin(.01t))$, which illustrates that this
sort of function actually has multiple periodic components
(the amplitudes of these different components can be evaluated
using Bessel functions, see Gray \& Mathews 1895). The above term in the 
disturbing function can therefore  be expressed as the  
following series:
\begin{equation}
\mathcal{R}'_{1,-1} \propto \sum_jC_j\sin(\varpi-\varpi_B+jLt-j\theta_L),
\end{equation}  
which is exactly the form which we are seeking (cf. Equation~\ref{eqho} above).
A single disturbing term of the form given in Equation~\ref{term1}
could therefore potentially  lead to the formation of multiple, evenly-spaced
features, like the gaps in the Cassini Division. 
\nocite{Gray1895}

Given all this, it seems reasonable to look for
three-body-like resonances in the Cassini Division where one of the
bodies is Mimas and the other is the B-ring edge. Furthermore, 
we will be particularly interested in terms in the disturbing function
of the B-ring edge that have the form:
\begin{equation}
\mathcal{R}_{1,-1}\propto d_1\tilde{d}_2\sin([\lambda-\varpi_B]-[2\lambda-2\lambda_{Mimas}-2\tilde{\phi}_L])
\end{equation}

\subsection{Evaluating the Disturbing Function}

The above analysis indicates that we will need two pieces
of the disturbing function: (1) The part of the disturbing
function due to Mimas that goes like 
$\cos(\lambda+\varpi-2\lambda_{Mimas})$ and (2)
the part of the disturbing function due to the B ring that
goes like $\cos(\lambda+\varpi_B-2\lambda_{Mimas}-2\tilde{\phi}_L)$.

The relevant part of the Mimas disturbing function is given by Equation~\ref{distfn} above:
\begin{equation}
\mathcal{R}_{Mimas}=\frac{Gm_M}{a_{M}}f(a/a_{M})e\cos(\lambda+\varpi-2\lambda_{Mimas})
\label{rmimas}
\end{equation}
where $m_M$ and $a_M$ are Mimas' mass and orbital semi-major axis, 
respectively, while $a$ and $e$ are the particles' semi-major axis and eccentricity, 
(recall that for a 2:1 resonance $f(\alpha)=-0.75/\alpha$). \nocite{MurrayDermott}

Of course, extracting the relevant bit of the ring's
disturbing function requires more effort.
Again, consider a mass ribbon whose position is described by the following equation:
\begin{equation}
r=r_o-\delta r=r_o-d_1\cos(\phi-\varpi_B)-\tilde{d}_2\cos(2\phi-\tilde{\phi}_{LM})
\end{equation}
where for simplicity of notation we have introduced the term:
$\tilde{\phi}_{LM}=2\lambda_{Mimas}+2\tilde{\phi}_L$.

The disturbing function for a small mass element $dm$
of this ring on a particle at radius $r'$ and longitude $\lambda$ is given by:
\begin{equation}
\mathcal{R}_{dm}=Gdm\left[\frac{1}{r'}\sum_{l=1}^{\infty}\left(\frac{r}{r'}\right)^lP_l(\cos\psi)
				-\frac{r'}{r^2}\cos\psi\right]
\end{equation}
where $\psi$ is the difference in longitudes between the mass element and 
the particle of interest, so $\phi=\psi+\lambda$.

The total disturbing function for the ring is then this expression integrated over
all $\psi$:
\begin{equation}
\mathcal{R}_{ring}=\frac{Gm_r}{2\pi}\int_o^{2\pi}\left[
	\frac{1}{r'}\sum_{l=1}^{\infty}\left(\frac{r}{r'}\right)^lP_l(\cos\psi)
				-\frac{r'}{r^2}\cos\psi\right]d\psi
\label{ringfull}
\end{equation}
where $m_r$ is the total mass of the ribbon, which for now is assumed to be uniformly
distributed in $\phi$.

The terms we are interested in here are proportional to $(\delta r)^2=(r-r_o)^2$
because:
\begin{equation}
(\delta r)^2 = 2d_1\tilde{d}_2\cos(\psi+\lambda-\varpi_B)\cos(2\psi+2\lambda-\tilde{\phi}_{LM})+....
\end{equation}
\begin{equation}
(\delta r)^2 = d_1\tilde{d}_2\cos(\psi+\lambda+\varpi_B-\tilde{\phi}_{LM})+....
\end{equation}
\begin{equation}
(\delta r)^2 = d_1\tilde{d}_2\cos(\lambda+\varpi_B-\tilde{\phi}_{LM})\cos\psi+....
\label{deltarexp}
\end{equation}
So this is the lowest order term in $d_1$, $\tilde{d}_2$ that contains
the desired frequency term.

Re-writing $\mathcal{R}_{ring}$ so that $\delta r$ is explicit, we find:

\begin{equation}
\mathcal{R}_{ring}=\frac{Gm_r}{2\pi r'}\int_o^{2\pi}
	\sum_{l=1}^{\infty}\left(\frac{r_o-\delta r}{r'}\right)^lP_l(\cos\psi)d\psi
				-\frac{Gm_rr'}{2\pi}\int_o^{2\pi}\frac{1}{(r_o-\delta r)^2}\cos\psi d\psi
\end{equation}

Expanding both terms and keeping only terms proportional to $(\delta r)^2$,we obtain:
\begin{equation}
\mathcal{R}'_{ring}=\frac{Gm_r}{2\pi r'r_o^2}
	\sum_{l=2}^{\infty}\frac{l(l-1)}{2}\left(\frac{r_o}{r'}\right)^{l}
	\left[\int_o^{2\pi}(\delta r)^2 P_l(\cos\psi)d\psi\right]
				-\frac{3Gm_rr'}{2\pi r_o^4}\int_o^{2\pi}(\delta r)^2\cos\psi d\psi
\end{equation}
While both parts of this function are non-zero, the second part is a single term while
(as we will see below) the first part is a slowly converging series. Thus at
this point we will drop the second term and only keep the parts of the first
term containing the element explicitly listed in Equation~\ref{deltarexp}:
\begin{equation}
\mathcal{R}'_{ring}\simeq\frac{Gm_rd_1\tilde{d}_2}{2\pi r'r_o^2}
	\cos(\lambda+\varpi_B-\tilde{\phi}_{LM})
	\sum_{l=2}^{\infty}\frac{l}{2}\left(\frac{r_o}{r'}\right)^{l}
	\left[\int_o^{2\pi} (l-1)\cos\psi P_l(\cos\psi)d\psi\right]
\end{equation}

\begin{figure}[tb]
\centerline{\resizebox{5in}{!}{\includegraphics{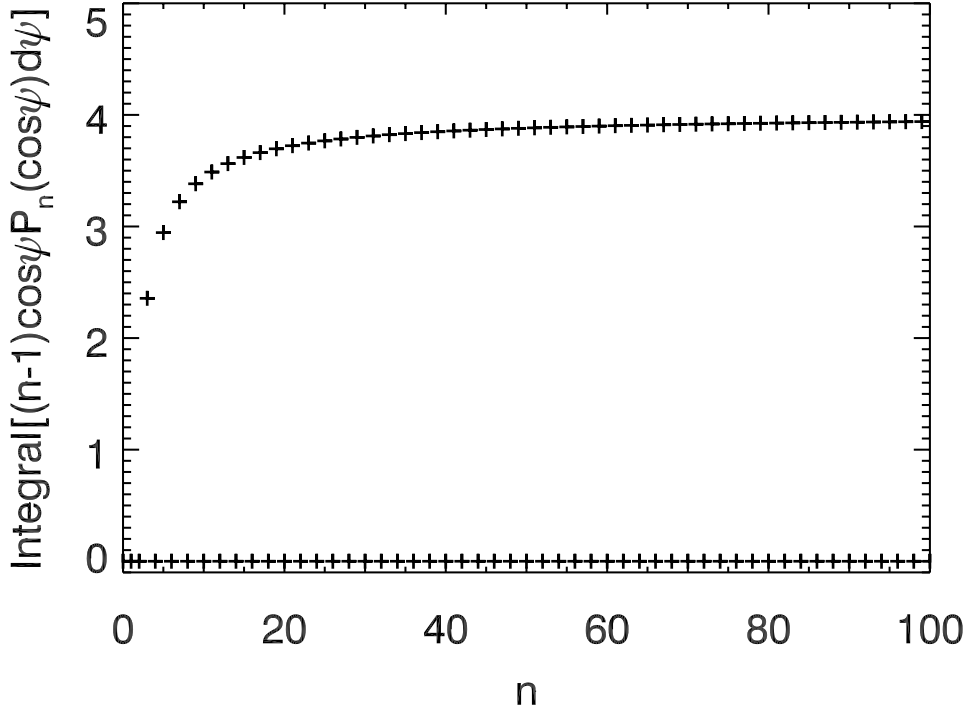}}}
\caption{The term $\int_o^{2\pi} (n-1)\cos\psi P_n(\cos\psi)d\psi$,evaluated
for different values of $n$. The bottom branch corresponds to even $n$, 
where the integral is zero by symmetry. The top branch corresponds
to odd $n$, which asymtotically approaches 4.}
\label{legendre}
\end{figure}

It turns out that for large values of $l$ the term in the square brackets 
becomes 4 for odd $l$ and 0 for even $l$ (see Fig.~\ref{legendre}). Since $r_o/r' \simeq 1$ in the 
Cassini division, the series slowly converges, so we can actually approximate this term as 2.
In this approximation:
\begin{equation}
\mathcal{R}'_{ring}\simeq \frac{Gm_rd_1\tilde{d}_2}{2\pi r'r_o^2}
	\cos(\lambda+\varpi_B-\tilde{\phi}_{LM})
	\sum_{l=2}^{\infty}l\left(\frac{r_o}{r'}\right)^{l},
\end{equation}
and since $\sum l x^l=1/(1-x)^2$
\begin{equation}
\mathcal{R}'_{ring}\simeq \frac{Gm_r}{2\pi r'}\frac{r'^2}{r_o^2}\frac{d_1\tilde{d}_2}{(r'-r_o)^2}
	\cos(\lambda+\varpi_B-\tilde{\phi}_{LM}).
\end{equation}

Finally, we assume in this case that the particle is on a nearly circular orbit, so $r'\simeq a$, so
the final expression for this function is:
\begin{equation}
\mathcal{R}^{r}_{ring}\simeq \frac{Gm_r}{2\pi a}\frac{a^2}{r_o^2}\frac{d_1\tilde{d}_2}{(a-r_o)^2}
	\cos(\lambda+\varpi_B-\tilde{\phi}_{LM})
\end{equation}
where the superscript $r$ indicates that this perturbation is due to the {\sl radial} 
distortions in the B ring edge.

This expression gives the disturbing function for a single ribbon of
mass total $m_r$. This perturbation from the entire outer part of the 
B ring can be approximated as a series of such ribbons with different 
values of $m_r, r_o, d_1$ and $\tilde{d}_2$. Since the structure of the
B ring is still uncertain at this point \citep{Hahn09}, for simplicity we will here 
assume that $d_1$ and $\tilde{d}_2$ decrease linearly with radius
towards the planet:
\begin{equation}
d_1=D_1\frac{r_o-R_i}{R_o-R_i}
\end {equation} 
\begin{equation}
\tilde{d}_2=\tilde{D}_2\frac{r_o-R_i}{R_o-R_i}
\end {equation} 
where $R_o$ is the mean radius of the B ring edge, $D_1$ and $\tilde{D}_2$
are the values of $d_1$ and $\tilde{d}_2$ at the edge, and $R_i=R_o-W$ is the
assumed radius where the amplitudes of the radial excursions in the mass
ribbons go to zero. Furthermore, let us assume that the mass of each ribbon
is:
\begin{equation}
m_r=2\pi\sigma r_o dr_o
\end{equation}
where $\sigma$ is the (unperturbed) surface mass density of the
outer B ring.

If we insert the above expressions for  $m_r, d_1$ and $\tilde{d}_2$ and
integrate over all values of $r_o$ between $R_i$ and $R_o$, 
assuming that $W=R_o-R_i$ is much less than $R_o$ and $a-R_o$, we find
the disturbing function from the entire B ring is given by:
\begin{equation}
\mathcal{R}^{r}_{Ring}\simeq \frac{G\sigma W}{3}\frac{a}{R_o}\frac{D_1\tilde{D}_2}{(a-R_o)^2}
	\cos(\lambda+\varpi_B-\tilde{\phi}_{LM})
\label{rring}
\end{equation}

At this point, we should note that the above model of the B ring might be too
simplistic, in that it assumes the mass ribbons are of constant mass.
It is of course possible that the mass density of the ribbons $\rho$ also varies with longitude:
\begin{equation}
\rho=\frac{m_r}{2\pi}[1+\mu_1\cos(\phi-\varpi_B)+\mu_2\cos(2\phi-\tilde{\phi}_{LM})],
\end{equation}
where $\mu_1$ and $\mu_2$ are fractional mass density variations along the ribbon.
If particles do not collide with one another, then the mass variations $\mu_1$ and
$\mu_2$ could be derived from the gradients in the radial variations and orbital
parameters. However, in practice the outer part of the B ring is very dense, 
so collisions are very important. Indeed preliminary investigations show that
the local optical depth of the B ring can vary by over a factor of two. While we
expect these density variations to have some connection with the radial positions
of the outer edge, at the present moment we will avoid attempting to estimate
$\mu_1$ and $\mu_2$. Regardless of the numerical values of these parameters, 
such mass anomalies would mean that terms of order $\mu_1\tilde{d}_2$ and $\mu_2d_1$
would also produce terms with the desired frequency. Following the same procedures
as described above, one can show the disturbing function for a ring with both radial
excursions and mass anomalies is (ignoring the possibility of phase
shifts between the mass anomalies and the radial excursions):

\begin{equation}
\mathcal{R}^{rm}_{Ring}\simeq \frac{G\sigma W}{3}\frac{a}{R_o}
	\left(\frac{D_1\tilde{D}_2}{(a-R_o)^2}-
	\mu_2\frac{D_1}{a-R_o}-\mu_1\frac{\tilde{D}_2}{a-R_o}\right)
	\cos(\lambda+\varpi_B-\tilde{\phi}_{LM})
\label{rmring}
\end{equation}

\subsection{Numerical Values of the Disturbing Function Terms.}

Having obtained these two terms, it is worth evaluating their
coefficients so we can make some quantitive comparisons of their strength:

For $\mathcal{R}_{Mimas}$ this is relatively easy, since:
\begin{equation}
\mathcal{R}_{Mimas}=\frac{Gm_M}{a_{M}}f(a/a_{M})e\cos(\lambda+\varpi-2\lambda_{Mimas})
\end{equation}
or, equivalently:
\begin{equation}
\mathcal{R}_{Mimas}=\frac{Gm_M}{a^2}\frac{a}{a_{M}}f(a/a_{M})(ae)\cos(\lambda+\varpi-2\lambda_{Mimas})
\end{equation}
Given $m_M \simeq 4*10^{19}$ kg, and $(a/a_M)f(a/a_M) \simeq -0.75$ for a 2:1 resonance, 
and assuming $a\simeq 118,500$ km and $ae \simeq 5$ km (appropriate for the Russell Gap), we find:
\begin{equation}
\mathcal{R}_{Mimas}=-7.1*10^{-4} m^2/s^2\left(\frac{ae}{5 km}\right)\left(\frac{118,500 km}{a}\right)^2\cos(\lambda+\varpi-2\lambda_{Mimas})
\end{equation}

For $\mathcal{R}_{Ring}$ there are more uncertain terms. First considering the
simpler case of pure radial excursions in the ring (Equation~\ref{rring}).
\begin{equation}
\mathcal{R}^{r}_{Ring}\simeq \frac{G\sigma W}{3}\frac{a}{R_o}\frac{D_1\tilde{D}_2}{(a-R_o)^2}
	\cos(\lambda+\varpi_B-\tilde{\phi}_{LM})
\end{equation}
Again, assume $a \simeq 118,500$ km for reference. Also, based on our measurements, 
we will assume $R_o \simeq 117,500$ km , $D_1 \simeq 20$ km and $\tilde{D}_2 \simeq 60$ km.
For rough purposes, we will use a surface mass density of $\sigma \simeq 100$ g/cm$^2$
and  $W \simeq 100$ km, though these parameters are less well constrained. Substituting these numbers
into the above expression gives the following result:

\tiny
\begin{equation}
\mathcal{R}^{r}_{Ring}\simeq 2.7*10^{-6} m^2/s^2*
\left(\frac{\sigma}{100 g/cm^2}\right)\left(\frac{W}{100 km}\right)
\left(\frac{a}{118,500 km}\right)
\left(\frac{D_1}{20 km}\right)\left(\frac{\tilde{D}_2}{60 km}\right)
\left(\frac{1000 km}{(a-R_o)}\right)^2
	\cos(\lambda+\varpi_B-\tilde{\phi}_{LM})
\end{equation}
\normalsize

Thus the perturbations from the B ring are about 200 times weaker than those from the
2:1 Mimas resonance. While this is a rather large factor, it is not many orders of magnitude.
Also note that the perturbations from the ring could be even larger
if there are significant variations in the mass of the ribbons with longitude.

\subsection{The Psuedo-Three-Body Resonance}

To complete the argument that the perturbations from the 
B ring and Mimas can act in concert to align the orbit pericenters
in the Cassini Division, we will now demonstrate the existence
of three-body-like resonances involving these objects. First,   
recall equation~\ref{eqho}, and realize that the
overall goal is to find a term in $d^2\varpi/dt^2$ that is
proportional to $\sin(\varpi-\varpi_B+jLt-\varphi_o)$. To do this, 
we need to express $d^2\varpi/dt^2$ in terms of the disturbing
function, which can be done by using the Lagrange planetary
equations.

Assuming the particles in the Cassini Division have no inclination, and keeping only
terms to lowest order in eccentricity $e$.
\begin{equation}
\frac{d\varpi}{dt}=\frac{1}{na^2e}\frac{\partial \mathcal{R}}{\partial e}+\dot{\varpi}_{sec},
\end{equation}
where $\dot{\varpi}_{sec}$ is again the precession induced by Saturn's oblateness, which
will we assume here to be a constant at any given semi-major axis $a$. This means that:
\begin{equation}
\frac{d^2\varpi}{dt^2}=\frac{d}{dt}\left(\frac{1}{na^2e}\frac{\partial \mathcal{R}}{\partial e}\right).
\end{equation}
In this case, we are only interested in the terms $\mathcal{R}_{Mimas}$ and
$\mathcal{R}_{Ring}$ derived above, and $\mathcal{R}_{Ring}$ does  
not depend on the eccentricity $e$, so this equation simplifies to:
\begin{equation}
\frac{d^2\varpi}{dt^2}=\frac{d}{dt}\left(\frac{1}{na^2e}\frac{\partial \mathcal{R}_{Mimas}}{\partial e}\right).
\end{equation}
Which can be expanded using Kepler's third law to give:
\begin{equation}
\frac{d^2\varpi}{dt^2}=
\frac{-1}{na^2e}\left[
\frac{1}{e}\frac{\partial \mathcal{R}_{Mimas}}{\partial e}\frac{de}{dt}+
\frac{1}{2a}\frac{\partial \mathcal{R}_{Mimas}}{\partial e}\frac{da}{dt}-
\frac{d}{dt}\left(\frac{\partial \mathcal{R}_{Mimas}}{\partial e}\right)\right].
\end{equation}
Since $de/dt$ and $da/dt$ can be expressed in terms of derivatives of 
the disturbing function, the first two terms in this expression naturally
give a product of two disturbing functions, enabling the term $\mathcal{R}_{Ring}$
to mix with  $\mathcal{R}_{Mimas}$ to obtain the desired frequency.
However, the third term can also do this since:
\begin{equation}
\frac{d}{dt}\left(\frac{\partial \mathcal{R}_{Mimas}}{\partial e}\right)=
\frac{\partial^2\mathcal{R}_{Mimas}}{\partial \epsilon\partial e}\frac{d\epsilon}{dt}+
\frac{\partial^2\mathcal{R}_{Mimas}}{\partial a\partial e}\frac{da}{dt}+....
\end{equation}
Note that other derivatives are not of interest here because either they are zero
or because they do not lead to terms with the appropriate frequencies.
For example, $\partial^2\mathcal{R}_{Mimas}/(\partial e)^2=0$, and any term
involving $d\varpi/dt$ cannot give a useful term because $\mathcal{R}_{Ring}$ is  
independent of $e$. Thus:
\begin{equation}
\frac{d^2\varpi}{dt^2}=
\frac{-1}{na^2e}\left[
\frac{1}{e}\frac{\partial \mathcal{R}_{Mimas}}{\partial e}\frac{de}{dt}+
\left(\frac{1}{2a}\frac{\partial \mathcal{R}_{Mimas}}{\partial e}
-\frac{\partial^2\mathcal{R}_{Mimas}}{\partial a\partial e}\right)\frac{da}{dt}-
\frac{\partial^2\mathcal{R}_{Mimas}}{\partial \epsilon\partial e}\frac{d\epsilon}{dt}
\right].
\end{equation}

Now we can use the following Lagrange equations:
\begin{equation}
\frac{da}{dt}=\frac{2}{na}\frac{\partial \mathcal{R}}{\partial \epsilon}
\end{equation}
\begin{equation}
\frac{de}{dt}=-\frac{e}{2na^2}\frac{\partial \mathcal{R}}{\partial \epsilon}
	-\frac{1}{na^2e}\frac{\partial \mathcal{R}}{\partial \varpi}
\end{equation}
\begin{equation}
\frac{d\epsilon}{dt}=-\frac{2}{na}\frac{\partial \mathcal{R}}{\partial a}
	+\frac{e}{2na^2}\frac{\partial \mathcal{R}}{\partial e}.
\end{equation}
Note that we want terms involving a product of $\mathcal{R}_{Mimas}$ and
$\mathcal{R}_{Ring}$, so only derivatives of the latter will be of interest here. 
Furthermore, since $\mathcal{R}_{Ring}$ only depends on $\lambda=nt+\epsilon$
and $a$, only the first term in each equation will contribute:

Hence, substituting and simplifying:
\begin{equation}
\frac{d^2\varpi}{dt^2}=
\frac{-1}{n^2a^4e}\left[
\left(\frac{1}{2}\frac{\partial \mathcal{R}_{Mimas}}{\partial e}-
\frac{a}{2}\frac{\partial^2\mathcal{R}_{Mimas}}{\partial a\partial e}\right)
	\frac{\partial\mathcal{R}_{Ring}}{\partial\epsilon}+
{2a}\frac{\partial^2\mathcal{R}_{Mimas}}{\partial \epsilon\partial e}
\frac{\partial\mathcal{R}_{Ring}}{\partial a}
\right].
\end{equation}
Notice that since $\mathcal{R}_{Mimas}$ depends on various powers of $a$, 
${\partial^2\mathcal{R}_{Mimas}}/{\partial a\partial e}$ is of order 
$a^{-1}\partial\mathcal{R}_{Mimas}/\partial e$, so all the elements of the first term are of the same order. However, since
$\mathcal{R}_{Ring} \propto (a-R_o)^{-x}$, $\partial\mathcal{R}_{Ring}/\partial a$
has terms that are of order $(a-R_o)^{-1}\mathcal{R}_{Ring}$, and since $a-R_o$
is much less than $a$, this means the latter term is much larger, so we can approximate:
\begin{equation}
\frac{d^2\varpi}{dt^2}\simeq
\frac{-2a}{n^2a^4e}
\frac{\partial^2\mathcal{R}_{Mimas}}{\partial \epsilon\partial e}
\frac{\partial\mathcal{R}_{Ring}}{\partial a}.
\end{equation}

Substituting in Equations~\ref{rmimas} and~\ref{rring} into this expression
(i.e., assuming $\mu_1=\mu_2=0$), we get:
\begin{equation}
\frac{d^2\varpi}{dt^2}=
\frac{2a}{n^2a^4e}
\left[\frac{Gm_M}{a_{M}}f(a/a_{M})\sin(\lambda+\varpi-2\lambda_{Mimas})\right]
\left[\frac{2G\sigma W}{3}\frac{a}{R_o}\frac{D_1\tilde{D}_2}{(a-R_o)^3}
	\cos(\lambda+\varpi_B-\tilde{\phi}_{LM})\right].
\end{equation}
Keeping only the desired long-period term, and recalling that $\tilde{\phi}_{LM} 
= 2\lambda_{Mimas}+2\tilde{\phi}_L$, this becomes:
\begin{equation}
\frac{d^2\varpi}{dt^2}=
\frac{2}{3}\frac{n^2}{e}\left[\frac{m_M}{M_S}\frac{a}{a_M}f(a/a_{M})\right]
\left[\frac{\sigma W R_o}{M_S}\frac{a^2}{R_o^2}\frac{aD_1\tilde{D}_2}{(a-R_o)^3}\right]
	\sin(\varpi-\varpi_B+2\tilde{\phi}_{L})
\end{equation}
Note that for a 2:1 resonance $a/a_Mf(a/a_M)\simeq-0.75$ (see Table 8.5
in Murray and Dermott 1999), so this equation can be written as:
\begin{equation}
\frac{d^2\varpi}{dt^2}=
-f_o^2\sin(\varpi-\varpi_B+2\tilde{\phi}_{L}).
\label{reseqho}
\end{equation}
where
\begin{equation}
f_o^2=
\frac{2}{3}\frac{n^2}{e}\left[\frac{m_M}{M_S}\frac{a}{a_M}|f(a/a_{M})|\right]
\left[\frac{\sigma W R_o}{M_S}\frac{a^2}{R_o^2}\frac{aD_1\tilde{D}_2}{(a-R_o)^3}\right].
\label{strength}
\end{equation}
is a positive quantity that indicates the strength of this resonance.

Recall that compound trigonometric functions like
$\sin(\varpi-\varpi_B+2\tilde{\phi}_{L})=\sin(\varpi-\varpi_B+2\phi_L\sin(Lt-\theta_L))$
can be expressed as a series of terms of the form $\sin(\varpi-\varpi_B+jLt)$.
We may therefore expand the right-hand side of Equation~\ref{reseqho} into the series of terms:
\begin{equation}
\frac{d^2\varpi}{dt^2}=
\sum_j-f_j^2\sin(\varpi-\varpi_B+jL t).
\label{reseqser}
\end{equation}
where $f_j^2=C_jf_o^2$ measures the strength of the individual resonances.
As shown in Figure~\ref{fftlib}, $C_j$ will in general be a decreasing function of
$j$, but for large-amplitude librations $C_j$ would not be much less than unity for 
most $j$ in the Cassini Division. 

If we consider the resonant argument $\varphi_j=\varpi-\varpi_B+jLt$, 
Equation~\ref{reseqser} says that there is a term in the equation of motion:
\begin{equation}
\ddot{\varphi_j}=-f_j^2\sin\varphi_j.
\end{equation}
Since $\ddot{\varphi_j}=\frac{d}{d\varphi_j}(\dot{\varphi_j}^2/2.)$, this means:
\begin{equation}
\dot{\varphi_j}^2=\dot{\varphi}_o^2-4f_j^2\sin^2(\varphi_j/2).
\end{equation}
 where $\dot{\varphi}_o$ is the value of $\dot{\varphi_j}$ where $\varphi_j$=0.
 Note a librating solution to this equation of motion will only
 exist if $\dot{\varphi_j}=0$ for some value of $\varphi_j$. This will only happen
 if $\dot{\varphi}_o^2<4f_j^2$. Or, in other words, when:
 \begin{equation}
 |\dot{\varpi}_o-\dot{\varpi}_B-jL|<2f_j
 \end{equation}
 where $\dot{\varpi}_o$ is the apsidal precession rate when $\varphi=0$. 
 Now, the apsidal precession rate in the Cassini Division is dominated by
 Saturn's oblateness. Thus this resonance
 will only be able to confine the pericenter locations over a region in the
 rings where the precession rate is within $2f_j$ of $\dot{\varpi}_B-jL$.
  Inserting the same numbers used in the previous section into Equation~\ref{strength}, 
we find that in the simple case without mass anomalies (and  the amplitude
of the $m=2$ pattern on the B ring edge is nearly constant):
\begin{equation}
f_j^2 \simeq7*10^{-22}/s^2 C_j\left(\frac{5 km}{ae}\right)
\left(\frac{\sigma}{100 g/cm^2}\right)\left( \frac{W}{100 km}\right)
\left(\frac{D_1}{20 km}\right)\left(\frac{\tilde{D}_2}{60 km}\right)
\left(\frac{1000 km}{a-R_o}\right)^3,
\end{equation}
so $2f_j$ is of order $5*10^{-11}$/s or 0.00025$^\circ$/day. 
Assuming the precession rates are dominated by Saturn's
oblateness (so $\delta\dot{\varpi}/\dot{\varpi}\simeq(7/2)\delta a/a$),
this range in precession rates corresponds to a range in
semi-major axis on the order of a few kilometers. 
These resonances should therefore be able to align the pericenters
of particle orbits over a region in the rings a few kilometers across,
which is comparable to the amplitudes of the observed eccentric 
features. This result is encouraging, although we must point out that
this analysis merely demonstrates that eccentric orbits within a few
kilometers of the resonance can have their pericenters aligned by
these resonances. This analysis does not determine
the actual eccentricity of individual particle orbits near these resonances,
which must be the subject of future work.

\section{Discussion of Potential Further Work}

The  above analysis shows that if the orientation of the 
B-ring edge librates at $\sim0.06^\circ$/day, then the combined 
gravitational perturbations from the B ring
and Mimas should align the pericenters of particle orbits
 at the observed locations of the eccentric
gap edges in the Cassini Division. Furthermore, if the
physical parameters of the B ring are close to the values 
assumed above, then particles with aligned orbital pericenters
will extend over a range of semi-major axis comparable to the observed
amplitudes of the eccentric features in the Cassini Division, which is needed
to avoid streamline crossing.
Both of these findings lend support to the notion
that the gaps in the Cassini Division may indeed be generated
by pseudo-three-body resonances involving perturbations
from the Mimas and the B-ring edge.  At the same time, 
it is also clear that more work needs to be done to
develop a complete model of how the Cassini-Division gaps
are formed in this sort of scenario. While such a
complete model is beyond the scope of this paper, 
we will describe in this section what we see as productive
avenues for future work  towards this goal.

First of all, while the above calculations provide an explanation
for why there should be coherent eccentric features at the observed
locations of the inner edges of the gaps in the Cassini Division, 
this analysis does not yet provide a physical
explanation of why gaps should form at these locations.
We expect that once the gaps have formed, the
relevant resonances should maintain the eccentric
shape of the inner edges by aligning the pericenters
of the relevant particles' orbits. Furthermore, we speculate
that the mass anomaly produced by the eccentric inner edge
of each gap may help keep the outer edge from diffusing inwards
via a modified shepherding mechanism in which the periodic radial and
tangential gravitational forces from the eccentric inner edge 
cause the semi-major axes of the particles on the gap's outer edge
to migrate outwards. 
While \citet{Rappaport98}  determined that the torque from a simple 
precessing eccentric ringlet is always negative and is therefore unable 
to confine the outer edge of a gap, the perturbations from eccentric
edges (which could involve significant mass anomalies as well
as well as radial excursions) have not yet been thoroughly investigated. 
Regardless of the specific shepherding mechanism,
any successful model of the gap outer edges must 
be consistent with the observed near-circularity of 
the outer edges. 

While further analytical calculations could clarify 
which torques might be able to hold open these gaps, 
numerical studies may be needed to determine how the
eccentric inner edges form in the first place. The above
calculations  suggest that the pericenters
of eccentric orbits will become aligned, but
they do not indicate whether there is any term in the 
equations of motion  that would
tend to generate eccentricities at these locations. 
While it is possible that such terms are produced
by some combination of the various perturbations
described above, it is also possible that the eccentricities
are generated by an instability in the Cassini Division 
itself. If particle collisions produce small eccentricities,
which are in turn aligned by the above resonance, this
will produce small anomalies in the local gravitational field
that could then drive larger eccentricities in the region,
which will also become aligned, until an eccentric edge is 
formed. If we can better understand how these
edges form, we may be able to convert the observed amplitudes
of the radial excursions into constraints on the
physical characteristics of the B ring such as
its mass density.

Whereas the amplitude of the edges' radial excursions 
should be related to the physical properties of the B-ring edge,
the orientations of the $m=1$ patterns should be related
simply to the kinematics of the B-ring edge. 
In the idealized model given above, the aligned
pericenter locations for the different edges
are determined by the movements of the 
different components in the B-ring edge, and
we expect this to be the case in the actual Cassini Division
as well. However, the actual pericenter locations
of the eccentric Cassini-Division features could
be affected by finite mass anomalies in the outer
B ring, as well as variations in the amplitude
of the $m=2$ pattern in the B ring edge. 
Given that these aspects of the B-ring
edge are not yet well modeled, we cannot yet
predict the absolute orientation
of the Cassini-Division edges at any given time.
In fact, if one accepts that the eccentric Cassini-Division 
edges are produced by the sorts of resonances described
above, one could even use the observed
amplitudes and orientations of the different eccentric edges
to place constraints on the  time variability in the B-ring's edge.

More generally, the above model was developed
assuming the $m=2$ pattern in the B-ring edge
librates with a frequency of around 0.06$^\circ$/day,
which matches the spacing in the observed pattern
speeds of the eccentric features in the Cassini Division.
More work needs to be done to confirm that this particular
solution is indeed the correct one. Beyond considering 
more data (e.g. from Cassini images, Spitale and Porco {\it in prep}), 
more sophisticated theoretical modeling of 
the B-ring edge may be able to better constrain the
likely solutions. Even if one of the other possible
solutions for the motion of the $m=2$ pattern
turns out to be correct, a variant of the above model
may still be able to produce the observed pattern
in the locations of the Cassini-Division features.
For example, if the $m=2$ pattern is actually circulating 
instead of librating, then its amplitude and drift rate
will periodically change as it drifts relative to Mimas, which should produce
a series of resonant terms in the equations analogous
to those derived above.

Finally, a full understanding of the 
Cassini Division gaps will also need to encompass the
noncircular ringlets in the Huygens, Herschel and Laplace gaps.
The Huygens ringlet, being the most eccentric
feature in the Cassini Division, may turn out
to be the limiting case of an eccentric edge.
The origin of the Laplace ringlet may 
be connected with the second-order
Pandora resonance on its inner edge \citep{Colwell09}. Even so, the
Herschel ringlet remains largely mysterious.

\section{Conclusions}

\begin{itemize}
\item 
The outer edges of five of the eight named gaps in the Cassini Division (that is, 
all the  gaps that do not contain dense ringlets) are circular to within 1 km.

\item
The inner edges of six of the eight gaps in the Cassini Division (that is, all of
the gap inner edges that do not lie near first-order Lindblad resonances with
known satellites) are eccentric, with $m=1$ radial variations that drift
around the planet at rates close to the expected apsidal precession rates.

\item 
The radial excursions in the  inner edge of the Barnard Gap, which lies close to the 5:4 ILR with Prometheus, do appear to have an $m=5$ component tied to that moon.

\item 
The radial excursions in the outer edge of the B-ring, which lies close to 
the 2:1 ILR with Mimas, have an $m=2$ component. However, the amplitude
of the pattern and its orientation relative to Mimas change with time. While the
available data do not yet determine a unique solution for the motion
of the $m=2$ pattern, one possibility
is that it librates relative to Mimas with a frequency $\sim0.06^\circ$/day and an amplitude
of $\sim 50^\circ$ in longitude (or a freqeuncy of $\sim0.12.^\circ$/day and an amplitude
of $\sim 30^\circ$).

\item
In addition to the $m=2$ pattern, the B-ring edge also appears to have an $m=1$ 
component that drifts around the planet at a rate close to the expected apsidal
precession  rate at the B-ring edge of $5.06^\circ$/day.

\item 
The pattern speeds of the eccentric features in the Cassini Division, including the
Huygens ringlet and the $m=1$ component of the B-ring edge, appear to form
a regular series given by $\Omega_p =\dot{\varpi}_B - jL$, where $j=0,1,2,3,...,7$, 
$\dot{\varpi}_B\simeq 5.06^\circ$/day is the apsidal precession rate at the B-ring edge, and 
$L\simeq 0.06^\circ$/day is a likely value for the libration frequency of the
$m=2$ component in the B-ring edge.

\item
By combining gravitational perturbations from both components of the B-ring edge
with the perturbations from the 2:1 Mimas ILR, one can find terms in the equations
of particle motion of the form $\ddot{\varpi} \propto \sin(\varpi-\varpi_B+jLt)$. Such terms could act to
align the pericenters of particle orbits at the locations of the eccentric inner edges
of each of the gaps, and therefore could explain the locations of the gaps
in the Cassini Division.

\end{itemize}

\section{Acknowledgments}

This work was carried out with financial support from NASA via the Cassini-Huygens program. 
We acknowledge the support of the Cassini Project and the VIMS team. We thank M. Evans, M, Tiscareno and R. French  for help in the development and validation of the code used to reconstruct
 the occultation geometries. We also thank the RSS team for sharing their data on the B-ring edge in advance of publication. We also thank J. Burns, J. Cuzzi, C. Murray, N. Rappaport, J. Spitale, and M. Tiscareno for stimulating and useful conversations.


\end{document}